\documentclass[a4paper,fleqn,usenatbib]{mnras}

\usepackage[T1]{fontenc}
\usepackage{ae,aecompl}

\usepackage{graphicx}	
\usepackage{amsmath}	
\usepackage{amssymb}	%
\usepackage{deluxetable}
\usepackage{xspace}
\usepackage{subfig}
\usepackage{nicefrac}

\usepackage{pdflscape}

\bibpunct{(}{)}{;}{a}{}{,}
\usepackage{hyperref}
\hypersetup{
    colorlinks,
    citecolor=blue,    filecolor=blue,
    linkcolor=blue,
    urlcolor=blue
}

\title[Satellites around the LMC]{The predicted luminous satellite populations around SMC and LMC-mass galaxies - A missing satellite problem around the LMC?}

\author[G. A. Dooley et al.]{\parbox{17cm}{
Gregory A. Dooley$^{1}$\thanks{e-mail: greg.dooley@gmail.com},
Annika H.G. Peter$^{2,3}$,
Jeffrey L. Carlin$^{4}$,
Anna Frebel$^{1}$,
Keith Bechtol$^{4}$,
and Beth Willman$^{5}$
}\vspace{0.3cm}\\
$^{1}$Department of Physics, Kavli Institute for Astrophysics and Space Research, Massachusetts Institute of Technology, \\ \ \ 77 Massachusetts Avenue, Cambridge, MA 02139, USA\\
$^{2}$CCAPP and Department of Physics, The Ohio State University, 191 W. Woodruff Ave., Columbus, OH 43210, USA \\
$^{3}$Department of Astronomy, The Ohio State University, 140 W. 18th Ave., Columbus OH 43210, USA\\
$^{4}$LSST, 933 North Cherry Avenue, Tucson, AZ 85721, USA \\
$^{5}$LSST and Steward Observatory, 933 North Cherry Avenue, Tucson, AZ 85721, USA\\
}

\date{Accepted by \mnras \ 2017 August 2. Received 2017 July 29; in original form 2017 March 5}

\pubyear{2017}

\newcommand{\msun}{$\, \mathrm{M_\odot}$\xspace}

\newcommand{\rvir}{$R_{\mathrm{vir}}$\xspace}
\newcommand{\mvir}{$M_{\mathrm{vir}}$\xspace}
\newcommand{\mtwo}{$M_{\mathrm{200}}$\xspace}
\newcommand{\mthree}{$M_{\mathrm{350}}$\xspace}
\newcommand{\mhalo}{$M_{\mathrm{halo}}$\xspace}
\newcommand{\minftwo}{$M_{\rm{200}}^{\rm{infall}}$\xspace}
\newcommand{\mpeakthree}{$M_{\rm{350}}^{\rm{peak}}$\xspace}
\newcommand{\mpeakvir}{$M_{\rm{vir}}^{\rm{peak}}$\xspace}
\newcommand{\mpeaktwo}{$M_{\rm{200}}^{\rm{peak}}$\xspace}

\newcommand{\mstarmhalo}{$M_* - M_{\rm{halo}}$\xspace}
\newcommand{\mstar}{$M_*$\xspace}

\newcommand{\vmax}{$v_{\rm{max}}$\xspace}
\newcommand{\msub}{$M_{\mathrm{sub}}$\xspace}
\newcommand{\mhost}{$M_{\mathrm{host}}$\xspace}
\newcommand{\vmaxpre}{$v_{\rm{max}}^{\rm{pre}}$\xspace}
\newcommand{\vmaxfilt}{$v_{\rm{max}}^{\rm{filt}}$\xspace}
\newcommand{\figpath}{}
\newcommand{\figwidth}{0.48}

\begin{document}
\label{firstpage}
\pagerange{\pageref{firstpage}--\pageref{lastpage}}
\maketitle

\begin{abstract}  
Recent discovery of many dwarf satellite galaxies in the direction of the Small and Large Magellanic Clouds (SMC and LMC) provokes questions of their origins, and what they can reveal about galaxy evolution theory.  Here, we predict the satellite stellar mass function of Magellanic Cloud-mass host galaxies using abundance matching and reionization models applied to the \textit{Caterpillar} simulations.  Specifically focusing on the volume within $50$~kpc of the LMC, we predict a mean of 4-8 satellites with stellar mass $M_* > 10^4$\msun, and 3-4 satellites with $80 < M_* \leq 3000$\msun. Surprisingly, all $12$ currently known satellite candidates have stellar masses of $80 < M_* \leq 3000$\msun.  Reconciling the dearth of large satellites and profusion of small satellites is challenging and may require a combination of a major modification of the \mstarmhalo relationship (steep, but with an abrupt flattening at $10^3$\msun), late reionization for the Local Group ($z_{\rm{reion}} \lesssim 9$ preferred), and/or strong tidal stripping. We can more robustly predict that $\sim 53\%$ of satellites within this volume were accreted together with the LMC and SMC, and $\sim 47\%$ were only ever Milky Way satellites. Observing satellites of isolated LMC-sized field galaxies is essential to placing the LMC in context, and to better constrain the \mstarmhalo relationship. Modeling known LMC-sized galaxies within $8$ Mpc, we predict 1-6 (2-12) satellites with $M_* > 10^5$\msun ($M_* > 10^4$\msun) within the virial volume of each, and 1-3 (1-7) within a single $1.5^{\circ}$ diameter field of view, making their discovery likely.
\newline
\end{abstract}


\section{Introduction}
The hierarchical structure formation predicted in Lambda Cold Dark Matter theories implies that dark matter halos of all scales contain substructure from past accretion events \citep{Springel08}. The Milky Way and M31 each fit this paradigm, both orbited by many known satellites. Moving down roughly an order of magnitude in total halo mass, the Large Magellanic Cloud should similarly contain its own substructure. \cite{Lynden-Bell76} made an initial proposition of the Large and Small Magellanic Clouds (LMC and SMC) as a dynamically linked group, followed by speculation that additional MW satellites and globular clusters were part of a greater Magellanic stream \citep{Lynden-Bell82, Lynden-Bell95}. In the years since, there has been a vigorous debate in the literature as to how many, if any, of the MW satellites were originally satellites of the LMC \citep{DOnghia08,Nichols11,Sales11}. 
One way to make progress is to probe the volume of space near the Clouds, especially the volume that remains tidally bound to them, for new satellites, as suggested by \citet{Sales11}.

Recently, surveys including Dark Energy Survey (DES), SMASH, Pan-STARRS, ATLAS, and MagLiteS have in fact revealed $\sim20$ new satellite candidates, many currently in the vicinity of the LMC \citep{Bechtol15, Drlica15, Kim15a, Kim15b, Koposov15, Laevens15, Martin15, Luque16, Torrealba16a, Drlica-Wagner16}. There is an ongoing debate as to which of these new satellites is physically associated with the Clouds.  \cite{Yozin15}, \cite{Koposov15} and \cite{Drlica15} each identify that the clustering of satellites near the LMC is highly improbable unless many satellites are dynamically associated with the Magellanic system. \cite{Deason15} considered satellites on an individual basis, comparing their positions and velocities to N-body simulations of LMC-sized galaxies and their substructure in a MW-sized system, and found that $2-4$ out of the $9$ then known DES satellites were likely part of the LMC. \cite{Jethwa16} followed with slightly different methodology, injecting LMC-sized galaxies into a MW potential with the known kinematics of the LMC, and found that $7-12$ of $14$ satellites were consistent with the LMC. However, their prediction relies on an assumption that the radial distribution of satellites follows that of all subhalos. If they were to instead use the more centrally concentrated radial distribution of luminous satellites (see \citet{Dooley16b}), their prediction would decrease by a few. \cite{Sales16} conducted a follow-up to the \cite{Sales11} study, and this time found $8$ of the $20$ new satellites possibly consistent with the LMC, with one, Horologium~I, definitely consistent. They agree with \cite{Koposov15b}, who had previously also concluded that Horologium~I is a satellite of the LMC. 

There remains much uncertainty over the fundamental theoretical question of how many satellites the LMC should have brought with it at infall. \cite{Jethwa16} predicts that a total of $70^{+30}_{-40}$ were accreted with the LMC and SMC. This means as many as $1/3$ of all MW satellites actually entered as part of the Magellanic Cloud system. In stark contrast, \cite{Sales16} predicts that only $5\%$ should have entered with the LMC, and \cite{Deason15} estimates a value of $7\%$, but allows for a range from a low of $1\%$ to a high of $25\%$.

It is hard to reconcile how many satellites should be observed around the LMC and SMC with how many are actually observed without a consistent and transparent prediction of satellite populations. As more satellites are discovered and/or confirmed, a prediction with clearly defined inputs can be tested to constrain the underlying galaxy formation physics in the model. The goal of this paper, a companion paper to \citet{Dooley16b}, is to show what simple, well-motivated theories predict for satellite populations in the LMC and SMC, as well as isolated galaxies of similar size to the LMC and SMC. Hence, we predict how many satellites were likely brought in with the LMC and SMC at infall, and how many satellites should exist within the local vicinity of the LMC today where many of the recent dwarfs were discovered. Any mention of the ``LMC'' by itself refers to the actual LMC. Any mention of ``LMC-size'', ``LMC-scale'', or ``LMC analog'' galaxies refers to galaxies which have a similar stellar mass to the LMC, but are not the LMC. The same applies to the SMC.

It is important to compare the LMC satellite population with those of isolated galaxies of comparable stellar mass, which provides a great opportunity to study satellite populations without the complicating dynamical issues of tidal stripping, ram pressure, and complex orbits of those which are in close proximity to the LMC, SMC, and MW. Additionally, the ambiguity of associating satellites with their original host does not exist for isolated galaxies. In that regard, isolated hosts are a cleaner sample through which observations can provide better empirical constraints on the estimates for the satellite population of our own LMC. These systems are also test grounds to study the effect of environment on dwarf galaxy star formation \citep{Wetzel15}, where gravitational effects are reduced and reionization may proceed differently than near the MW and M31. On the other hand, isolated galaxies are too distant to detect the extremely low luminosity satellites that are being found near the Magellanic Clouds. Thus to fully understand the satellite populations of LMC-sized hosts as well as to test our model for populating hosts with satellites in general, it is important to observe and analyze both LMC-size isolated galaxies and the LMC itself.

The prospects for finding dwarf galaxies around nearby LMC analogs in and beyond the Local Group is good, as new surveys prove that even ultrafaint dwarf galaxies can be identified at distances of $\sim$~a few Mpc. Recent surveys of resolved stellar populations in $\sim$Milky Way-mass halos are revealing faint dwarf satellites and their remnants (e.g., Cen~A: \citealt{Crnojevic14, Crnojevic16}; M~81: \citealt{Chiboucas13}; NGC~253: \citealt{Sand14,Romanowsky16}; NGC~891: \citealt{Mouhcine10}). There are even some isolated examples of dwarfs around dwarfs. For instance, the dwarf galaxy Antlia~B ($M_{\rm V} \sim -9.7$) is located near the Local Group galaxy NGC~3109, which has a stellar mass similar to the SMC \citep{Sand15}. The tidally disrupting dwarf galaxy NGC~4449B and its associated tidal stream (\citealt{MartinezDelgado12,Rich12,Toloba16}) has been discovered near NGC~4449 ($D = 4.3$ Mpc) which has a stellar mass similar to the LMC. Finally, the $M_{\rm V} = -7.7$ ($M_* \sim 10^5$\msun) dwarf MADCASH~J074238+652501-dw has been found around the $M_*\sim2 \times M_*^{LMC}$ host galaxy NGC~2403 ($D = 3.2$~Mpc; \citealt{Carlin16}). To date, systematic searches for satellite companions of lower-mass hosts are lacking.  However, the available capabilities of wide-field imagers on large aperture telescopes to resolve stellar populations to nearly the outer reaches of the Local Volume ($D \lesssim 8$~Mpc) are beginning to be exploited (e.g., \citealt{Carlin16}) to study the halos of LMC analogs beyond the Local Group.

Here, we provide the theoretical context for both the LMC satellite population and Magellanic Cloud analogs. In Section~\ref{sec:methods_lmc}, we describe our theoretical approach.  In Section~\ref{sec:results_lmc}, we present our main results. First, we highlight some surprising results when we compare simple theoretical models of the satellite population around the LMC to the observed population. Namely, we show that the vicinity of the LMC has a statistically significant dearth of satellites with $10^4$\msun$< M_* < 10^7$\msun, especially compared with the number of satellites with $M_* \lesssim 10^3$\msun. This primary result is seen in Fig.~\ref{fig:LMC50kpc}. We present different hypotheses to explain this discrepancy. Finally, we make predictions for satellite abundances around specific target galaxies over a range of host galaxy stellar masses near that of the LMC that are located between $2$ and $8$ Mpc from the MW. Observations of these systems can help unravel the ``missing satellite'' puzzle of why so few (currently zero) satellites with $10^4$\msun$< M_* < 10^7$\msun are found within $50$~kpc of the LMC. To guide observers, we further provide estimates on the radial dependence within a line of sight of satellite abundances to motivate and compare to observational searches. In Section~\ref{sec:conclusions}, we summarize our key findings and make both theory- and observation-oriented recommendations to solve this puzzle of the dwarf satellite population near the LMC.

\section{Methods}
\label{sec:methods_lmc}
We apply the same techniques used in \cite{Dooley16b} to make predictions for the distribution of possible satellites around host galaxies. Given the stellar mass of a host galaxy, we generate a random realization of satellites around it according to the following procedure:
\begin{enumerate}
\item
Convert the host's stellar mass to a total halo mass (see section 2.5 of \cite{Dooley16b} for details).
\item
Determine the typical subhalo mass function (SHMF) for host galaxies.
\item
Sample the SHMF to generate a random realization of dark matter subhalos for the host.
\item
Model reionization by assigning subhalos to be dark or luminous according to a probability function that depends on a halo's peak or infall mass.
\item
Assign a stellar mass to each luminous subhalo according to an abundance matching (AM) model.
\item
\textit{(Optionally)} Assign distances to each satellite according to a radial distribution profile of luminous satellites.
\end{enumerate}
We repeat this procedure $30,000$ times for each calculation of interest in order to obtain a convergence of the mean and to sample the variance.

The SHMF, fraction of galaxies that survive reionization, radial distribution of satellites, and infall distribution of satellites are found by analyzing simulated galaxies from the \textit{Caterpillar} simulation suite \citep{Griffen16}. This consists of $33$ high particle resolution ($m_{\rm{p}} = 3\times 10^4$\msun) and high temporal resolution ($320$ snapshots each) zoom-in simulations of Milky Way-sized galaxies.

As in \cite{Dooley16b}, we model reionization's ability to leave dark matter halos entirely dark as follows: a halo's maximum circular velocity must reach a critical value, \vmaxpre, before the redshift of reionization, $z_{\rm{reion}}$, or reach a larger critical value after reionization, \vmaxfilt, in order to form stars. Applying this model to the \textit{Caterpillar} suite leads to a function indicating the fraction of halos that are luminous as a function of halo mass, as seen in fig. 3 of \cite{Dooley16b}. We choose values to replicate the model of \cite{Barber14} with $z_{\rm{reion}} = 13.3$, finding \vmaxpre$=9.5 \, \rm{km/s}$ and \vmaxfilt$=23.5 \, \rm{km/s}$. Unless otherwise stated, we use this as our ``baseline'' reionization model.

A reionization redshift of $z = 13.3$ is relatively early, especially compared to the value of $7.8 \leq z_{\rm{reion}} \leq 8.8$ estimated by \cite{Planck16b}. However, in order to remain consistent with the model used in \cite{Dooley16b}, which leads to a prediction for the number of MW satellites with $M_* > 10^3$\msun consistent with that of completeness corrected observations \citep{Hargis14}, we stick with $z_{\rm{reion}} = 13.3$ in our baseline reionization model. We then also explore models where reionization occurs later for comparison. Furthermore, the physical meaning of $z_{\rm{reion}}$ in our model is imprecise since decreasing it is degenerate with decreasing \vmaxpre. Decreasing either parameter increases the fraction of halos which form stars.

Due to large differences in AM models, particularly at low stellar masses, we implement a total of five AM models. These models are the Brook model \citep{Brook14}, the Moster model \citep{Moster13}, the GK14 model \citep{GarrisonKimmel14}, the GK16 model \citep{GarrisonKimmel16}, and the Behroozi model \citep{Behroozi13}.

We define the halo virial radius, \rvir, using the \cite{Bryan98} fitting function for the radius at which a halo is virialized in the spherical tophat model. At $z=0$ for our cosmological parameters, \rvir is the radius such that the mean enclosed halo density is $104$ times the critical density of the universe, $\rho_c = 3H_0^2/8 \rm{\pi} G$. \mvir refers to the gravitationally bound mass within \rvir, and any mention of $R_{\mathrm{\Delta}}$ or $M_{\mathrm{\Delta}}$ refers to the radius and mass of a halo where the mean enclosed density is $\rm{\Delta}$ times the critical density.

All of our steps are identical to those in \cite{Dooley16b} except for two adjustments. First, we use a slightly different SHMF. We find that a single function in which abundances are directly proportional to the host halo mass is sufficient to describe the SHMF over a host mass interval of 1 dex, but begins to become less accurate outside the range it was calibrated to. Since the total halo mass of the SMC and LMC is around 1 dex smaller than the MW mass, we calibrate the mass function specifically to halos in a mass range that encompasses the estimated masses of the LMC and SMC. We select a subset of isolated field halos in the simulations (outside the virial radius of the MW-like host and within the contamination radius) which have masses in the range $5 \times 10^{10} < M_{\rm{vir}} < 5 \times 10^{11}$\msun. The differential abundance of subhalos follows the form
\begin{equation}
\label{eq:SHMF_lmc}
\frac{\rm{d}N}{\rm{d}M_{\rm{sub}}} = K_0 \left(\frac{M_{\rm{sub}}}{\rm{M_{\sun}}} \right)^{-\alpha_{\rm{mf}}} \frac{M_{\rm{host}}}{\rm{M_{\sun}}},
\end{equation}
where the mass definition of \mhost and \msub vary according to AM model. The total number of subhalos counted, however, always refers to the number within the \cite{Bryan98} virial radius. The function is controlled by two fit parameters: the logarithmic slope $\alpha_{\rm{mf}}$ and normalization factor $K_0$. The values we obtain and use are $\alpha_{\rm{mf}} = 1.93$, $K_0 = 0.00588$ for \msub$=$\mpeakvir and \mhost$=$\mvir which is needed for the GK14 and GK16 models, $\alpha_{\rm{mf}} = 1.88$, $K_0 = 0.00219$ for \msub$=$\minftwo and \mhost$=$\mtwo which is needed for the Moster model, and $\alpha_{\rm{mf}} = 1.88$, $K_0 = 0.00282$ for \msub$=$\mpeakthree and \mhost$=$\mthree which is needed for the Brook model. These values lead to a $\sim 20\%$ reduction in abundance predictions compared to the values used for the smaller dwarf field halos in \cite{Dooley16b}. For predictions of the MW, we use the same SHMF parameters as listed in \cite{Dooley16b}.

Second, the LMC is massive enough such that we can no longer assume the subhalo abundances are Poisson distributed. As pointed out in \cite{BoylanKolchin10}, and confirmed in \cite{Mao15} and \cite{Lu16}, subhalo abundances more accurately follow a negative binomial distribution where the variance in the mean number of satellites increases relative to that of a Poisson distribution as $M_{\rm{sub}} /M_{\rm{host}}$ decreases. More quantitatively, the variance of $N(> M_{\rm{su}b}/M_{\rm{host}})$ is $\sigma^2 = \sigma_P^2 + \sigma_I^2$, where $\sigma_P^2 = \langle N \rangle$ is the Poisson variance, and $\sigma_I^2 = s_I^2 \langle N \rangle^2$ is an intrinsic scatter. The value of the fractional intrinsic scatter, $s_I$, was found to be $0.18$ in \cite{BoylanKolchin10}. Studying the variance in \textit{Caterpillar} halos with our own mass functions, we find $s_I = 0.14$ to be a better fit. We implement this by generating random samples of dark matter halos in logarithmic bins of $M_{\rm{sub}} /M_{\rm{host}}$. The number of halos per bin is chosen according to a negative binomial distribution whose variance increases for lower mass ratios. For our scale of galaxies, the variance in the number of satellites with $M_* > 10^3$\msun is larger than that of a Poisson distribution, but is roughly the same for satellites with $M_* > 10^5$\msun.

\section{Results}
\label{sec:results_lmc}
In the next two subsections, we consider two applications of our methods. First, we predict the satellite galaxy distribution within the vicinity of the LMC itself, and compare it to currently known dwarf galaxies in that region. To do this, we approximate the LMC and SMC as once-isolated galaxies that accreted on to the MW. Second, we make predictions for isolated galaxies with comparable stellar mass to the LMC and SMC. This allows us to estimate and tabulate the number of satellites that may be found in distant galaxies of similar size. The two investigations are mutually-reinforcing in that improved understanding of one can lead to improved modelling and predictions for the other.

Both subsections require an estimate of the total halo mass of the LMC and SMC if they were isolated at $z=0$, as well as the virial radius of each. We assume stellar mass values of $2.6 \times 10^9$\msun for the LMC and $7.1 \times 10^8$\msun for the SMC. These are calculated, as are all stellar masses of Local Volume galaxies in this work, based on the stellar luminosity which is derived using the $K_{\rm S}$-band magnitude and distance from \citealt{Karachentsev13}\footnote{Updated version available at \url{http://www.sao.ru/lv/lvgdb/}.}, assuming $M_{\rm Ks, \odot} = 3.28$, and $M/L = 1$; see also \citealt{vandermarel02,Harris09}. From these stellar masses, we infer the total halo mass as in \cite{Dooley16b}, using the AM models with an assumed $0.2$ dex $1 \sigma$ log-normal scatter in the \mstarmhalo relationship. The GK16 model is an exception, in which its scatter is larger and increases towards lower mass halos. For the LMC, we find a value of \mvir$= 2.3 \times 10^{11}$\msun for the GK models, $2.1 \times 10^{11}$\msun for the Behroozi model, $1.7 \times 10^{11}$\msun for the Moster model, and $1.8 \times 10^{11}$\msun for the Brook model. This corresponds to a virial radius of $156, 153, 141$ and $146 \, \rm{kpc}$. For the SMC, we find \mvir$= 1.3 \times 10^{11}$\msun for the GK models, $1.1 \times 10^{11}$\msun for the Behroozi model, $9.2 \times 10^{10}$\msun for the Moster model, and $9.5 \times 10^{10}$\msun for the Brook model. The corresponding radii are $132, 123, 116,$ and $117 \, \rm{kpc}$.

The masses we infer for the LMC are slightly lower than the value of \mtwo$= 2-2.5 \times 10^{11}$\msun (which converts approximately to \mvir$= 2.3-3.9 \times 10^{11}$\msun) estimated by \cite{Nichols11}, and \cite{Penarrubia16}, but larger than the LMC analog mass of $3.6 \times 10^{10}$\msun used in \cite{Sales16} and slightly larger than the upper limit from \cite{Jethwa16}.

\subsection{Number of LMC and SMC satellites}
\label{sec:satellite_counts_lmc}
In this section, we consider the actual LMC and SMC, quantifying the number of their satellites. We want to estimate the number of satellites they each had upon accretion onto the MW, which corresponds to the number of present-day MW satellites which were once, or still are, satellites of the Magellanic Clouds. To do so, we must estimate their halo mass at infall, which we infer from their stellar mass. However, estimating their stellar masses at infall accurately is difficult due to uncertainty over their infall time. Depending on the bulk velocity of the LMC, the mass of the MW and the Magellanic Clouds, the evolving MW gravitational potential, models for the Magellanic Stream, and LMC-SMC interactions, the accretion time of the LMC could be as recent as $1$ Gyr ago \citep{Busha11b}, longer than $4$ Gyr ago \citep{Bekki11}, or anywhere from $1 - 12$ Gyr ago \citep{Shattow09, Besla07,Kallivayalil13}.

Fortunately, this very large uncertainty does not significantly impact the number of satellites we predict. If the LMC accreted onto the MW $5$ Gyr ago for instance, its stellar mass at infall would be $\sim 30\%$ less than today according to the LMC star formation history computed in \cite{Harris09}. A $30\%$ decrease in the stellar mass of an LMC-sized host only decreases the predicted abundance of satellites by $13\%$ (as seen in Fig.~\ref{fig:ngreater_lmc}). When limiting predictions to fixed volumes rather than the virial radius, the difference drops yet again. Within $50$ kpc, the change is $8\%$. Moreover, uncertainty in the present day stellar mass of the LMC and SMC is large, at $~ -30\%/+70\%$ according to \cite{Harris09}. With no obvious best solution for identifying the stellar mass at infall, we simply use the present day estimate of stellar mass for the LMC and SMC, acknowledging that this uncertainty leads to a $\sim\pm 15\%$ uncertainty in satellite abundances in our model.

\begin{figure}
\centering
\includegraphics[width=\figwidth\textwidth]{\figpath 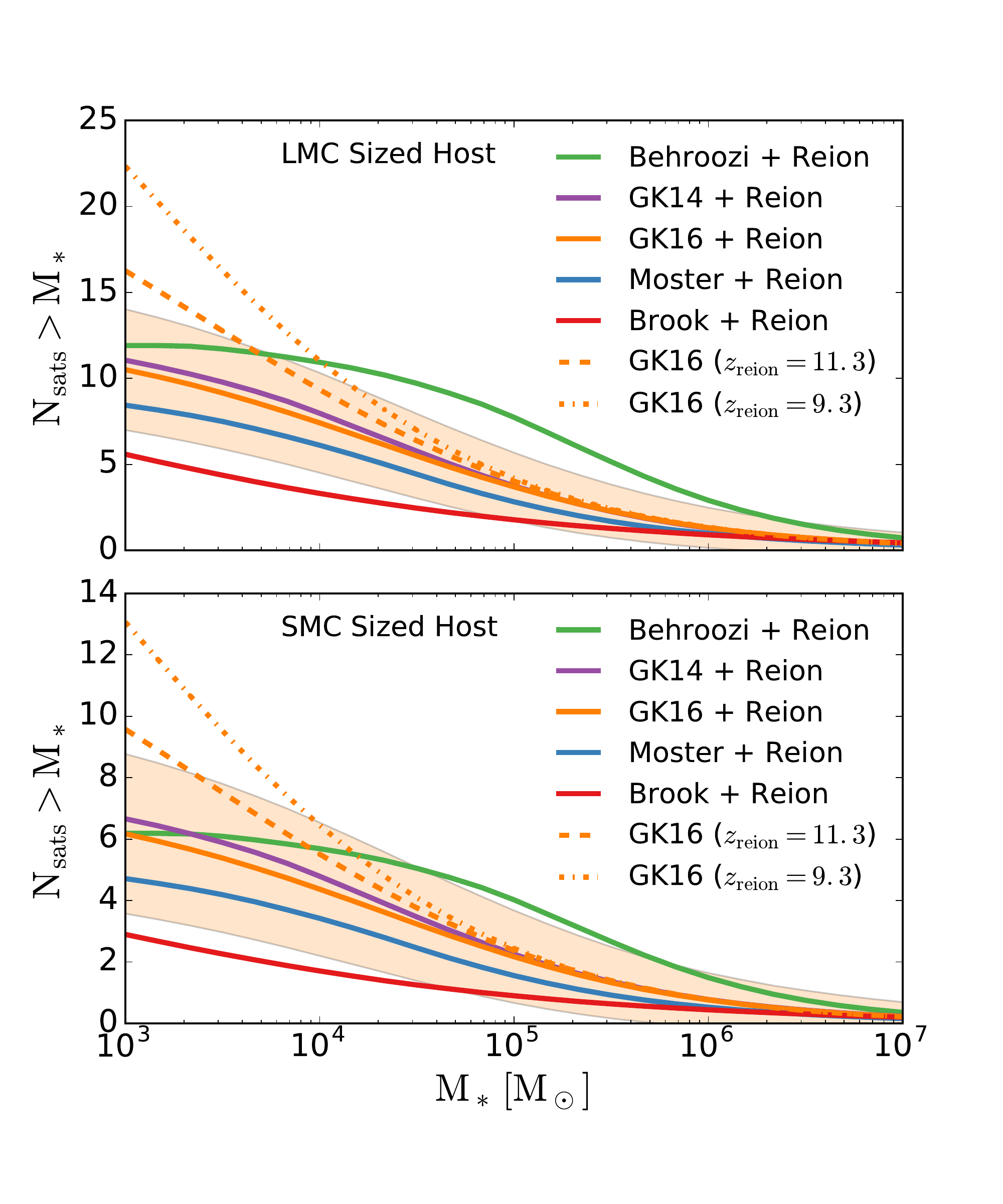}
\caption[Predicted satellite abundance around an isolated LMC- and SMC-sized hosts]{\emph{Upper Panel: }Mean number of satellites around an isolated LMC-sized host galaxy as a function of the minimum satellite stellar mass. \emph{Lower Panel: }Mean number of satellites around an isolated SMC-sized host as a function of the minimum satellite stellar mass. The shaded band in each panel shows the one $\sigma$ variation in satellite abundance for the GK16 model. The effect of delaying reionization by shifting its starting redshift in our model from $z=13.3$ (solid lines) to $z=11.3$ (dashed orange line) and $z=9.3$ (dot-dashed orange line) is shown for the GK16 model. Reionization controls the fraction of low-mass galaxies which can form stars.}
\label{fig:ngreater_minlum_lmc}
\end{figure}

In Fig.~\ref{fig:ngreater_minlum_lmc}, we show our prediction for the number of satellites within an isolated halo's virial radius as a function of the minimum satellite stellar mass. We do this for hosts with stellar masses of the LMC and SMC. Particularly for low stellar masses of the satellites, the number of satellites predicted varies greatly due to different abundance matching models, super Poissonian noise, and uncertainty in reionization. Regarding AM models, most of the discrepancy between the GK models and the Moster model arises from their different predictions of total halo mass and virial radius, with the Moster model predicting smaller values. The Brook model assigns lower stellar masses to halos than the rest, so much so that its predictions for UFD MW satellites are below the completeness limit estimates of \cite{Hargis14} and \cite{Drlica15} as shown in \cite{Dooley16b}. In that regard it can be considered a lower limit abundance matching model for the number of satellites with $M_* > 10^3$\msun. The Behroozi model, on the other hand, is known to overpredict the number of MW satellites with $M_* > 10^5$\msun \citep{GarrisonKimmel14, Dooley16b}, and can be seen as an upper limit in that mass range.

We highlight uncertainty due to counting statistics and reionization on the GK16 model since it is calibrated down to a lower \mstar than the other models, is the only model to explicitly model scatter in the \mstarmhalo relationship, and is the most recent model. The shaded orange band indicates the $\pm 1 \sigma$ range due to counting statistics about the mean. With the GK16 model and our baseline reionization, we predict that $7 - 14$ and $4-9$ satellites with $M_* > 10^3$\msun are accreted into the Milky Way with the LMC and SMC respectively. For satellites with $M_* > 10^5$\msun, we predict $2-6$ and $1-4$. 

Reionization in our baseline model occurs relatively early, at $z=13.3$. We demonstrate how delaying reionization affects the abundance of faint satellites by including predictions for the GK16 model with reionization occurring at $z=11.3$ and $z=9.3$. When reionization begins later, it suppresses fewer low mass halos leading to a large increase in galaxies with $M_* < 10^5$\msun. The effect is diminished for larger galaxies, with almost no change occurring for satellites with $M_* > 10^5$\msun. More observations of low mass satellites are needed to provide better joint AM and reionization model constraints.

Extrapolated down to $\sim 10^2$\msun to compare with \citet{Jethwa16}, all of our models predict significantly fewer than their $70^{+30}_{-40}$ satellites associated with LMC and SMC combined unless reionization only suppresses star formation in halos of \mpeakvir$\lesssim 10^8$\msun, consistent with the later work of \citet{Jethwa16b}.  If we extrapolate the GK16 model with $z_{\rm{reion}} = 9.3$ down to 10$^2$\msun in stars, which is the model in Fig. \ref{fig:ngreater_minlum_lmc} with the closest prediction to the \citet{Jethwa16} numbers at $52.5 \pm 8.5$ satellites, we find too many satellites for the Milky Way given by the completeness-corrected estimates of \citet{Hargis14} and discussed in \citet{Dooley16b}.  \citet{Jethwa16} fix the functional form of the LMC and SMC luminosity functions while allowing the normalization to float.  A consistent subhalo abundance matching model between the Clouds and the Milky Way is not enforced in their work, while it is in ours.  We attribute the discrepancy between our result and that of \citet{Jethwa16} to this fact.

\subsubsection{Stellar mass function of LMC vicinity satellites}
\label{sec:lmcvicinity}
Fig.~\ref{fig:ngreater_minlum_lmc} predicts the number of satellites around LMC and SMC-sized galaxies if they were isolated. However, the actual LMC and SMC are located only $50$ and $60$ kpc from the MW Galactic Center. Using a MW mass of $1.4 \times 10^{12}$\msun, and approximating the dark matter halo density distributions as Hernquist profiles \citep{Hernquist90}, we estimate their tidal radii as the distance to the $L_3$ Lagrange point and get $15-17$ kpc for each. Thus nearly all of the former Magellanic satellites would now no longer be gravitationally bound to either the LMC or the SMC. Even if not bound though, former Magellanic satellites that accreted close to the LMC and SMC are likely still spatially correlated with them. This is especially true if the Magellanic Clouds are on their first pericentric passage, as favored by \cite{Besla07, Busha11b, Kallivayalil13} and \cite{Sales16}, although they could be on second or third passages \citep{Shattow09, Bekki11}. 

Since most of the recently discovered satellites near the LMC are within $50$ kpc of it \citep{Bechtol15, Drlica15}, and in particular all $12$ satellites that \cite{Jethwa16} found to have a probability of $>50 \%$ to be associated with the LMC are within $50$ kpc, we focus our predictions on that subvolume. Beyond $50$ kpc, satellites are more influenced by tides from the MW and are more mixed with MW satellites. We approximate that the number of Magellanic satellites within $50$ kpc of the LMC and SMC at infall is still the same number within those distances presently. In this regard, our predictions are an upper limit, and are more accurate the more recently the LMC and SMC accreted on to the MW since their satellites would have less time to migrate away. We comment on the effect of satellites migrating away from the LMC and SMC later in our analysis. 

We choose this approach, rather than a full dynamical treatment of \cite{Jethwa16} because \textit{Caterpillar} has few LMC analogs with suitable halo mass and kinematics, and no strong analog of the LMC/SMC pair \cite{Griffen16}. Our approach also gains us the ability to precisely place a hypothetical LMC and SMC in their known spatial position relative to the MW. We refrain from modelling a specific survey volume since most of the volume within $50$ kpc of the LMC has been surveyed by DES and MagLiteS, to the extent that we do not expect more than approximately one as-yet undetected satellite to exist here which is brighter than the first MagLiteS candidate. Moreover, ongoing surveys will make comparisons to the $50$ kpc radius volume increasingly accurate.

The expected number of satellites in our target subvolume of a $50$ kpc radius sphere centered on the LMC comes from three sources: the MW, the LMC, and the SMC. Assuming a reionization model, a MW mass of $1.4 \times 10^{12}$\msun, and an isotropic satellite distribution, we integrate the radial distribution of satellites from each source over the LMC subvolume, placing the MW at $50$ kpc from the LMC, and the SMC at $24$ kpc from the LMC. The radial distribution of satellites is determined from the \textit{Caterpillar} simulations, selecting only subhalos which we identify as luminous in our baseline reionization model. This is a very critical step, since the radial distribution of luminous satellites is much more concentrated than that of all dark matter subhalos, one of the biases discussed in \cite{Sawala16}. We would under predict the combined satellite abundance near the LMC by a factor of $\sim 2.5$ if we used the radial distribution of all $z=0$ subhalos.

We find that the normalized radial distribution of satellites within the host halo virial radius does not depend on the mass of the host. The distribution does depend weakly on subhalo peak mass, with satellites having \mpeakvir$>10^9$\msun being less centrally concentrated than those with \mpeakvir$<10^9$\msun, which are subject to more selection effects from reionization due to being smaller. However, the difference between these distributions leads to only a $10\%$ change in our predictions for the LMC, which is within the $1 \sigma$ uncertainty of the mean of our radial distribution. We therefore use a single radial distribution as a function of $r/$\rvir calibrated to all luminous satellites. We tested for this on all isolated halos within the \textit{Caterpillar} suite with a mass of $10^{10} < M_{\rm{vir}} < 3 \times 10^{12}$\msun.  The functional form of our radial distribution, and a plot comparing the distribution of luminous satellites against all subhalos, can be found in \cite{Dooley16b}.

\begin{figure}
\centering
\includegraphics[width=\figwidth\textwidth]{\figpath 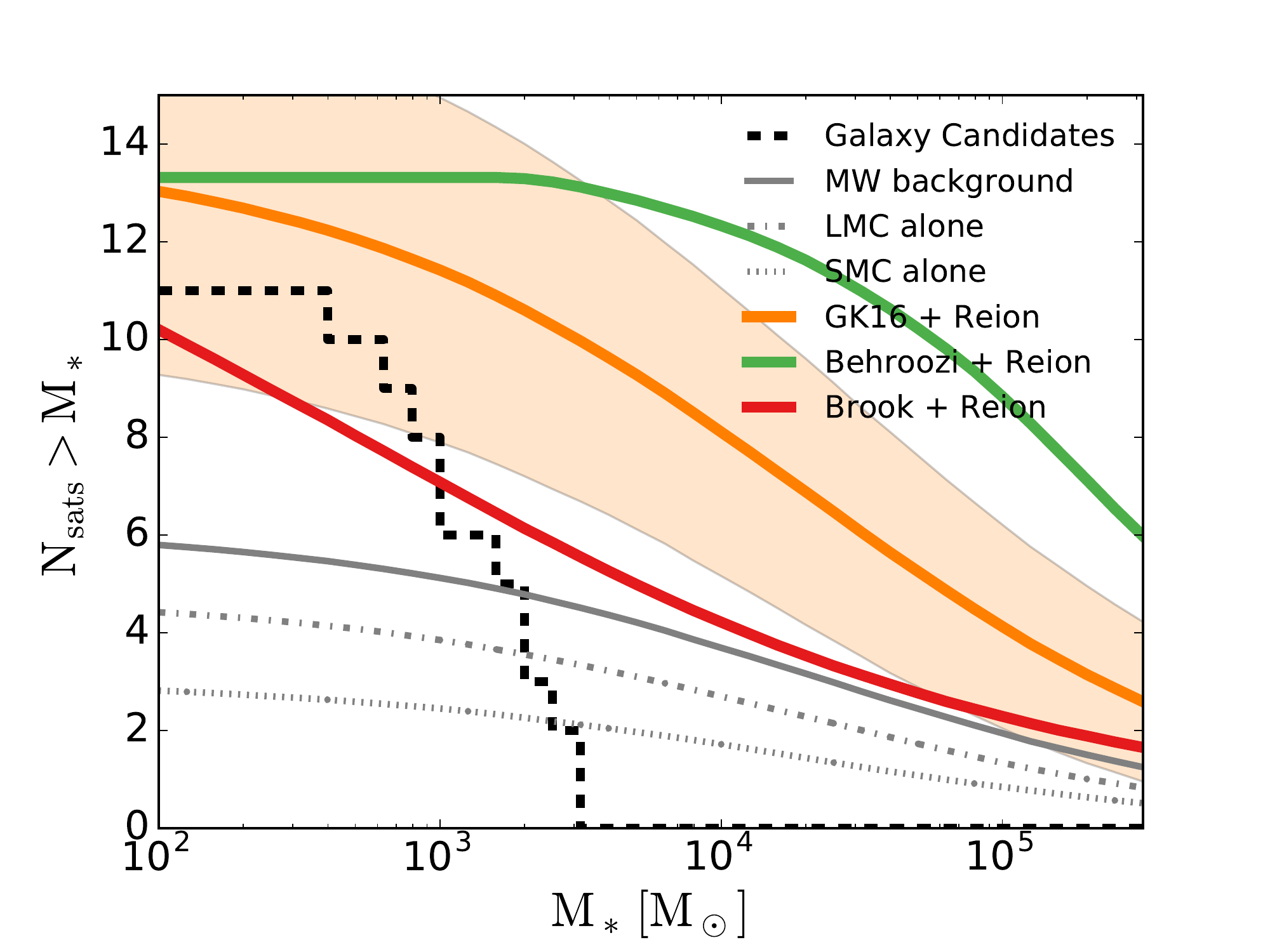}
\caption[Predicted and actual satellite abundance within $50$ kpc of the LMC]{Predicted and actual number of satellites within $50$ kpc of the LMC as function of minimum stellar mass. Dashed black line shows the stellar mass function of all currently known candidate galaxies. At an estimated stellar mass of just $82$\msun, Cetus~II falls off the mass range plotted. For galaxies with $M_* < 10^4$\msun, this can be regarded as a lower limit since more low stellar mass galaxies are expected to be found in ongoing surveys. Also plotted are predictions for the stellar mass function according to the Behroozi, GK16, and Brook AM models. For the GK16 model in orange, we include shaded bands indicating $1 \sigma$ variance due to counting statistics and uncertainty in the radial distribution. Predictions from the GK14 and Moster models are nearly identical to the GK16 model and therefore not shown. The GK16 prediction is broken down into contributions of satellites from the MW background, the LMC, and the SMC.} 
\label{fig:LMC50kpc}
\end{figure}

In Fig.~\ref{fig:LMC50kpc}, we show the predicted cumulative satellite stellar mass function for galaxies in the LMC subvolume. We show this for the Behroozi, GK16, and Brook models. Due to restricting our predictions to a small fixed volume, the Moster, GK14, and GK16 models all predict nearly identical values, so we show just the GK16 model for simplicity. As the most recent and versatile model, we split the predictions for the GK16 model into contributing components from the MW, LMC, and SMC, as shown in grey. The sum of all three components is shown by the solid orange line, with $1 \sigma$ variation due to counting statistics shown by the shaded orange band.

We compare these predictions to the dashed black line, which shows the cumulative stellar mass function for all currently known satellite candidates within $50$ kpc of the LMC. A major caveat is that this function is likely a lower limit for satellites with $M_* < 10^4$\msun, since the full extent of the volume we are considering has not yet been surveyed and analyzed for such faint galaxies \citep{Drlica-Wagner16}. Not included in the observed sample are the LMC and SMC themselves, since we purposefully choose our small subvolume to contain these two rare galaxies. Whether we include Magellanic Cloud sized systems in our predictions or not is unimportant, since so few satellites of their size are predicted in all models.

The catalog of MW satellites which contribute to the dashed black line in  Fig.~\ref{fig:LMC50kpc}, including their names, stellar masses, and distances from the LMC, are tabulated in Table~\ref{table:LMCsats}. The data for all MW satellites were taken from the compilation of \cite{McConnachie12},\footnote{Available online at \url{http://www.astro.uvic.ca/~alan/Nearby_Dwarf_Database.html}.} which we supplemented with the following systems discovered more recently than the last update of the McConnachie catalog in Sept 2015: Virgo I \citep{Homma16}, Pictor II \citep{Drlica-Wagner16}, Crater 2 \citep{Torrealba16b}, Aquarius 2 \citep{Torrealba16a}, and DES~J0225+0304 \citep{Luque16}.
Stellar masses were calculated based on the derived $V$-band absolute magnitudes, doubling the luminosity to account for $M_{\rm V}$ being the luminosity within the half-light radius, and assuming a stellar mass to light ratio of one. Systematic uncertainty in the luminosity of satellites up to a factor of two makes little difference in the interpretation and qualitative nature of Fig.~\ref{fig:LMC50kpc}. A factor of $10$ increase in luminosity would be needed to significantly alter the interpretation.

\begin{table}
\tablewidth{0.48\textwidth}
\centering
\caption{Satellite galaxies within $50 \, \rm{kpc}$ of the LMC}
\label{table:LMCsats}
\begin{tabular}{lcc} 
\hline
\hline
\textbf{Name} &  \boldmath$M_* \ \rm{[M_\odot]}$ & \boldmath$d_{\rm{LMC}} \ \rm{[kpc]}$  \\ 
\hline
Cet~II*  & $82$  &  46 \\  
Tuc~V*  & $370$ &  30 \\  
Eri~III*   & $520$ & 48 \\  
Tuc~III* & $760$ & 33 \\   
Hor~II* & $870$  &  38 \\  
Ret~II & $990$  & 24  \\  
Pic~II* & $1600$  & 12 \\  
Ret~III* &  $1700$  & 44  \\  
Hor~I  & $1900$  & 38 \\   
Tuc~IV* & $2100 $ & 27 \\  
Tuc~II  & $2700$ & 37 \\  
Gru~II*  &  $3000$ &  46 \\  
\hline
\multicolumn{3}{l}{$^{*}$ not yet spectroscopically confirmed}
\end{tabular}
\end{table}

After selecting only those within $50$~kpc of the LMC, our sample (in order of smallest to largest distance from the LMC) consists of Pic~II, Ret~II, Tuc~IV, Tuc~V, Tuc~III, Tuc~II, Hor~I, Hor~II, Ret~III, Cet~II, Gru~II, and Eri~III. Of these, only three have been spectroscopically confirmed as dwarf galaxies: Reticulum~II \citep{Koposov15b,Simon15,Walker15}, Horologium~I \citep{Koposov15b} and Tucana~II \citep{Walker16}. This means some candidates in our sample may end up not being true dwarf galaxies. For instance, spectroscopic follow-up of Tucana~III suggests it is a tidally stripped dwarf galaxy, but its status as a galaxy is not definitive \citep{Simon16}. Ongoing spectroscopic analysis, however, suggests that most, if not all candidates, will be confirmed as galaxies \citep{FermiLAT16}. Uncertainty in satellite positions could mean gaining $0-2$ satellites with $M_* < 10^3$\msun, which would not be enough to change the interpretation nor qualitative nature of Fig.~\ref{fig:LMC50kpc}.


More galaxies are also expected to be present, since this population comes from an incomplete survey area around the LMC. The Magellanic Satellites Survey (MagLiteS) is likely to uncover more ultrafaint satellites near the LMC in a footprint not already surveyed by DES \citep{Drlica-Wagner16}. It is also possible that within the DES footprint, faint dwarf galaxies (those with $M_* < 10^4$\msun) will continue to be discovered. Between the likelihood that candidate galaxies will be confirmed as real galaxies, and the expectation of discovering more galaxies, the observed stellar mass function in Fig.~\ref{fig:LMC50kpc} should be considered a lower limit.

Even with these caveats, there is a dramatic disagreement between the predicted and observed stellar mass functions. All AM models greatly overestimate the number of satellites with $M_* > 3 \times 10^3$\msun, or more importantly, with $M_* > 10^4$\msun, a mass range where few, if any, new satellites are likely to be discovered. Following assumptions in \cite{Bechtol15}, a galaxy with stellar mass of $M_* = 10^4$\msun corresponds to $M_{\rm V} \sim -5.1$, which in the year 1 DES survey would be expected to be found with high efficiency out to $300 \, \rm{kpc}$. To remain undetected, a satellite of this size would need a half-light radius larger than $\sim 300 \, \rm{pc}$, making the surface brightness comparable to or lower than that of the lowest surface brightness MW satellites \citep{Torrealba16b}. Since the DES survey covered approximately half of the volume within $50$~kpc of the LMC and found zero $M_* > 10^4$\msun satellites \citep{Drlica-Wagner16}, it is unlikely the remaining volume contains more than two such satellites. 

In contrast to zero observed $M_* > 10^4$\msun satellites, the Brook model predicts $\sim 4$, the GK16 model $\sim 8$, and the Behroozi model $\sim 12$. Additionally, all models predict far too few satellites in the interval $10^2 < M_* < 3 \times 10^3$\msun. Once again, the Behroozi model is in particularly strong disagreement, a fact not surprising in light of its shortcomings in predicting satellite galaxies already discussed in \cite{GarrisonKimmel14} and \cite{Dooley16b}. Quantifying the disagreement in terms of random chance, the GK16 model predicts $7.9 \pm 3.0$ satellites with $M_* > 10^4$, and a $0.04\%$ chance of zero galaxies. For the Brook model, odds are improved, but only to $1.5\%$.

There is some evidence that this subvolume happens to have fewer satellites with $M_* > 10^4$\msun by chance. A total of $8$ such satellites (excluding the LMC and SMC) are known within the MW out to a galactic distance of $100$ kpc. If they were isotropically distributed and follow the radial distribution from our model, $1.4$ satellites would be expected in the $50$ kpc radius volume under consideration. An additional easing of tensions could be made if the Canis Major overdensity ($M_* \approx 4.5 \times 10^7$\msun) is in fact a dwarf galaxy since it is within $50$ kpc of the LMC. However, too much contention exists regarding whether it is \citep[e.g.,][]{Martin04,Bellazzini04,MartinezDelgado05,Bellazzini06,deJong07}, or is not (e.g., \citealt{LopezCorredoira06, Moitinho06, Momany06, RochaPinto06}; for a recent summary of this debate, see \citealt{Yanny16}) a galaxy to include it in our sample.

\subsubsection{Reconciling theory and observations?}
\label{sec:solutions}
Since we are analyzing a range of satellites far fainter than those used to calibrate any AM model, it is perhaps not surprising that our predictions do not agree well with observations. We therefore explore a range of possible alterations to our model which improve alignment with the data, and assess the plausibility of each. 

To make discussions easier, we define satellites with $M_* > 10^4$ as ``large UFDs,'' and satellites with $10^2 < M_* < 3 \times 10^3$ as ``small UFDs.'' In these terms, the problem with our predicted satellite abundance is a matter of too many large UFDs, and too few small UFDs. 

There are several ways in which the number of predicted large UFDs can be reduced which we identify as important to consider:
\begin{enumerate}
\item
Lowering the mass of the MW, LMC and/or SMC.
\item
Original LMC and SMC satellites have migrated to larger distances.
\item
Less centrally concentrated radial distribution of satellites.
\item
Tidal stripping.
\item
Steeper \mstarmhalo relationship.
\end{enumerate}

Each of these options simultaneously exacerbate the problem of predicting too few small UFDs. The predicted number of small UFDs can be increased without substantially increasing the predicted number of large UFDs in the following three ways:
\begin{enumerate}
\item
Reionization occurs later.
\item
The halo size threshold needed for star formation before reionization is reduced.
\item
The \mstarmhalo relationship deviates from a power law. It is ``bent'' near a stellar mass of $10^3$\msun to have a more flat slope.
\end{enumerate}

\subsubsection*{Reduced MW/LMC/SMC Mass}
Any reduction in the mass used to model the MW, LMC, or SMC would decrease the predicted number of satellites at all stellar mass scales. However, this can not by itself be an explanation for the over-prediction of large UFDs. In the extreme event that the MW, LMC, and SMC are all one-half the total halo mass we use, the number of large UFDs predicted for the LMC subvolume in the GK16 model decreases $30\%$ from $7.9$ to $5.6$, still a large statistical discrepancy from zero.

\subsubsection*{LMC/SMC Satellite Migration}
We have so far assumed that the number of satellites within $50$ kpc of the LMC and SMC at infall is equal to the number within $50$ kpc today. Especially if the LMC/SMC pair is on its second or third pericentric passage around the MW, the positions of the original Magellanic satellites would be changed by the MW gravitational potential, becoming less concentrated in the immediate vicinity of the LMC and SMC. This would reduce the predicted number of LMC and SMC satellites, but would once again not be sufficient by itself to explain the over-prediction of large UFDs. If the number of LMC and SMC satellites within $50$ kpc of the LMC are reduced by one half due to satellite migration, the baseline GK16 model would still predict $5.7$ large UFDs.

\subsubsection*{Radial Distribution}
A radial distribution that is less centrally concentrated would reduce the predicted contribution of satellites from all three of the MW, LMC, and SMC. The uncertainty in the mean of the normalized radial distribution of luminous satellites that we determine from the \textit{Caterpillar} halos leads to a $1 \sigma$ uncertainty of $28\%$ in the number of satellites within $50$ kpc of the LMC, regardless of satellite mass range. A reduction by this amount changes the GK16 prediction of large UFDs from $7.9$ to $5.7$, still too high to be statistically consistent with zero.

\subsubsection*{Tidal Stripping}
Tidal stripping can reduce the stellar mass of predicted large UFDs, helping to alleviate tension with observations by shifting the predicted stellar mass function curve to the left. The fact that most UFDs fall on the same mass-metallicity relationship as larger galaxies \citep{Kirby13b} suggests that most UFDs are not tidally stripped. However, there is evidence of possibly significant tidal stripping in the UFDs Segue~II \citep{Kirby13}, Hercules \citep{Roderick15, Kupper16}, Leo~V \citep{Collins16}, and Tucana~III \citep{Drlica15, Simon16}. Of those, only Tucana~III is in the sample of satellites within $50$ kpc of the LMC. However, most of the satellites in the sample have not yet been studied in follow-up campaigns to determine what levels of tidal stripping they may have undergone. It is therefore plausible that more satellites will show evidence of tidal stripping.

\begin{figure}
\centering
\includegraphics[width=\figwidth\textwidth]{\figpath 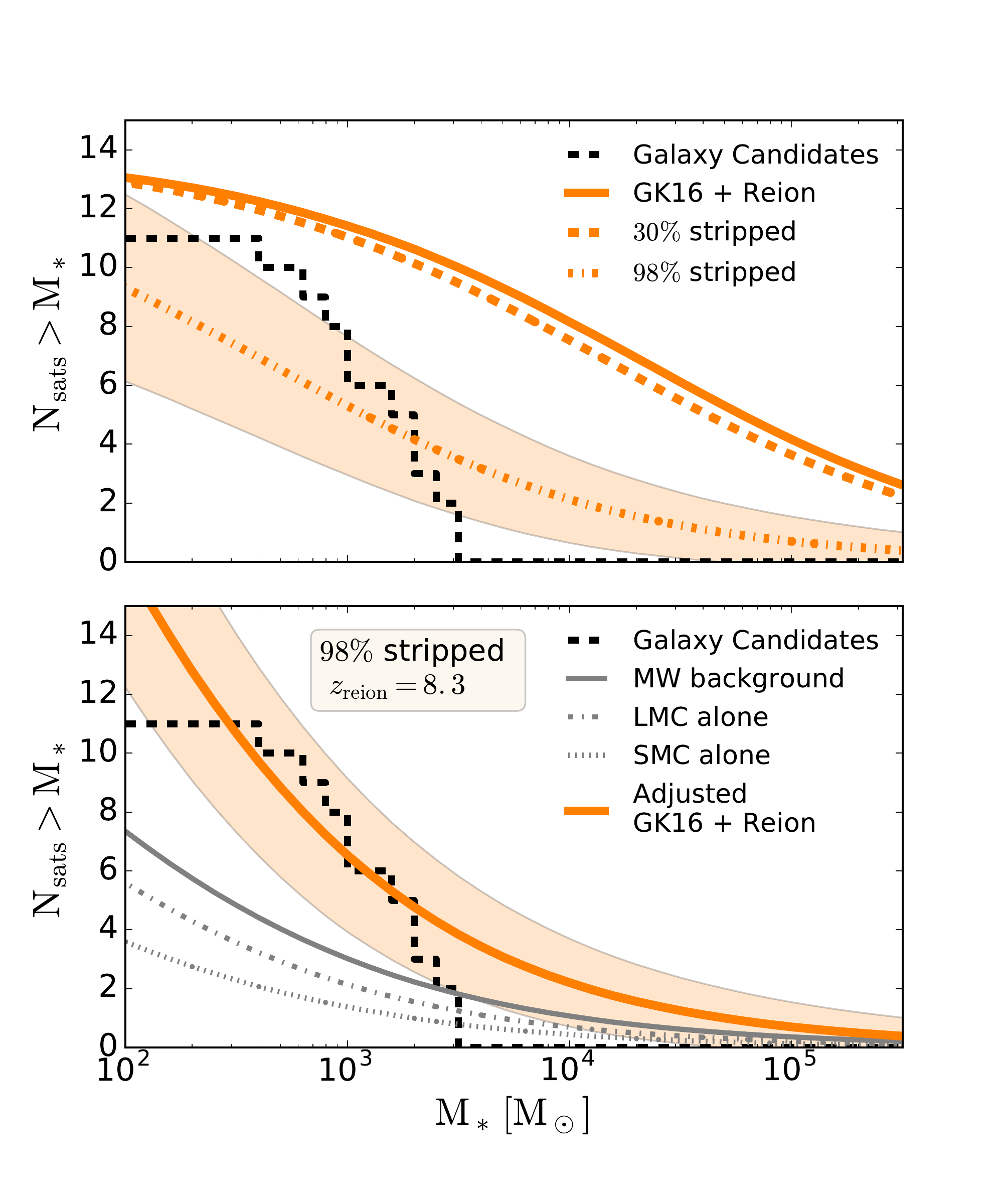}
\caption[Predictions for the LMC vicinity satellite stellar mass function with tidal stripping]{\emph{Upper panel: }Predictions for the stellar mass function of satellites within $50$ kpc of the LMC according to the GK16 model with $0\%$, $30\%$, and $98\%$ of stellar mass stripped from satellites. Only with extreme amounts of stripping are the number of predicted satellites with $M_* > 10^4$\msun diminished enough to be within $2\sigma$ of the number of observed satellites, zero. One sigma variation is shown around the $98\%$ stripped model with a shaded band. \emph{Lower panel: }Same prediction of the GK16 model with $98\%$ of stellar mass stripped except with reionization shifted from $z=13.3$ to $z=8.3$.  
A similar result could be achieved by lowering the halo size threshold needed for the first galaxies to form.}
\label{fig:tidallystripped}
\end{figure}

In the top panel of Fig.~\ref{fig:tidallystripped} we show the original GK16 stellar mass function prediction and the prediction if all satellites are stripped. Only a massive amount of tidal stripping, $>95\%$ of stellar mass stripped, could explain the excess of large UFDs by itself. This would reduce the number of large UFDs from $\sim 8$ to $\sim 2$. Using particle tagging in the \textit{Caterpillar} simulations, we conduct a quick estimate of the fraction of stars that could be stripped. We select satellites which form stars before reionization and survive to $z=0$ with the stellar mass of UFDs. We then tag their $2\%$ most bound particles at the time of peak mass, and assume no star formation proceeds thereafter. This is roughly consistent with their classification as reionization fossils, in which $> 70\%$ of their stars are estimated to have formed before reionization \citep{Bovill09, Brown12, Brown14a, Brown14b}. The satellites which are approximately $50$ kpc from the MW at $z=0$, corresponding to the distance from the observed LMC vicinity satellites to the MW, have on average $30\%$ of their stars stripped. We include a model of $30\%$ stellar mass stripping in the top panel of Fig.~\ref{fig:tidallystripped} and find it has little quantitative effect on the predicted number of satellites at any mass scale. 

Our tagging scheme ignores the complexities of satellite orbits and additional tides due to the LMC/SMC system, and the gravitational potential of the MW disk, both of which would increase the fraction of stellar mass stripped.  Additionally, the subhalos in our simulations have cuspy density profiles, whereas if UFD satellites are cored (for instance due to self-interacting dark matter \citep{Dooley16}, or to stellar feedback \citep{Governato10,Read16}) they would be more susceptible to tidal stripping \citep{Penarrubia10,Zolotov12, Brooks13}. Even so, it is unlikely that tidal stripping can be the main driver of the unusual shape of the stellar mass function near the LMC.  We find it highly implausible that $\geq 95\%$ of stellar mass is stripped on average in the LMC vicinity satellite galaxies without evidence of more tidal tails or strong deviation from the mass-metallicity relation. Moreover, a galaxy that has been stripped so much is unlikely to be identified as such \cite{Wetzel10}.  We expect that the presence of a stellar disk in either the MW or the Clouds will be indiscriminate in their stripping with respect to satellite mass \cite{GarrisonKimmel17}, even if the core size is mass-dependent.  In \citet{Dooley16}, we found that the presence of large, self-interacting dark matter cores (larger than typical feedback-driven cores in CDM) only modestly affected the survival rate of UFD satellites relative to cusped satellite halos.  Because cores encompass only a small percent of the total mass of halos, the halo (and hence, the luminous part of the satellite) must be heavily stripped before the difference in halo profile leads to divergent tidal stripping.  Thus, tidal stripping is unlikely to explain the dearth of large ultrafaints and abundance of small ultrafaints.

\subsubsection*{Steeper \mstarmhalo relationship}
Lastly, we consider a steeper \mstarmhalo relationship as part of the solution to over-predicting large UFDs. For stellar masses with $M_* < 10^8$\msun, $M_* \propto M_{\rm{halo}}^\alpha$ in all AM models considered. Already from comparing the Brook AM model ($\alpha = 3.1$) to the GK16 model ($\alpha = 1.97$) in Fig.~\ref{fig:LMC50kpc}, it is apparent that a steeper \mstarmhalo relationship reduces the predicted number of low stellar mass satellites. We can easily study this using the GK16 model because it allows for a range of possible \mstarmhalo slopes that all are consistent with the classical dwarf scale MW and M31 satellites. In the model, a steeper slope is paired with higher scatter in the \mstarmhalo relationship so that the reduction of classical dwarf-sized galaxies due to the steeper slope is compensated by an increase in the number of low mass dark matter halos that upscatter to contain higher stellar mass galaxies than the mean of the relationship. 

\begin{figure}
\centering
\includegraphics[width=\figwidth\textwidth]{\figpath 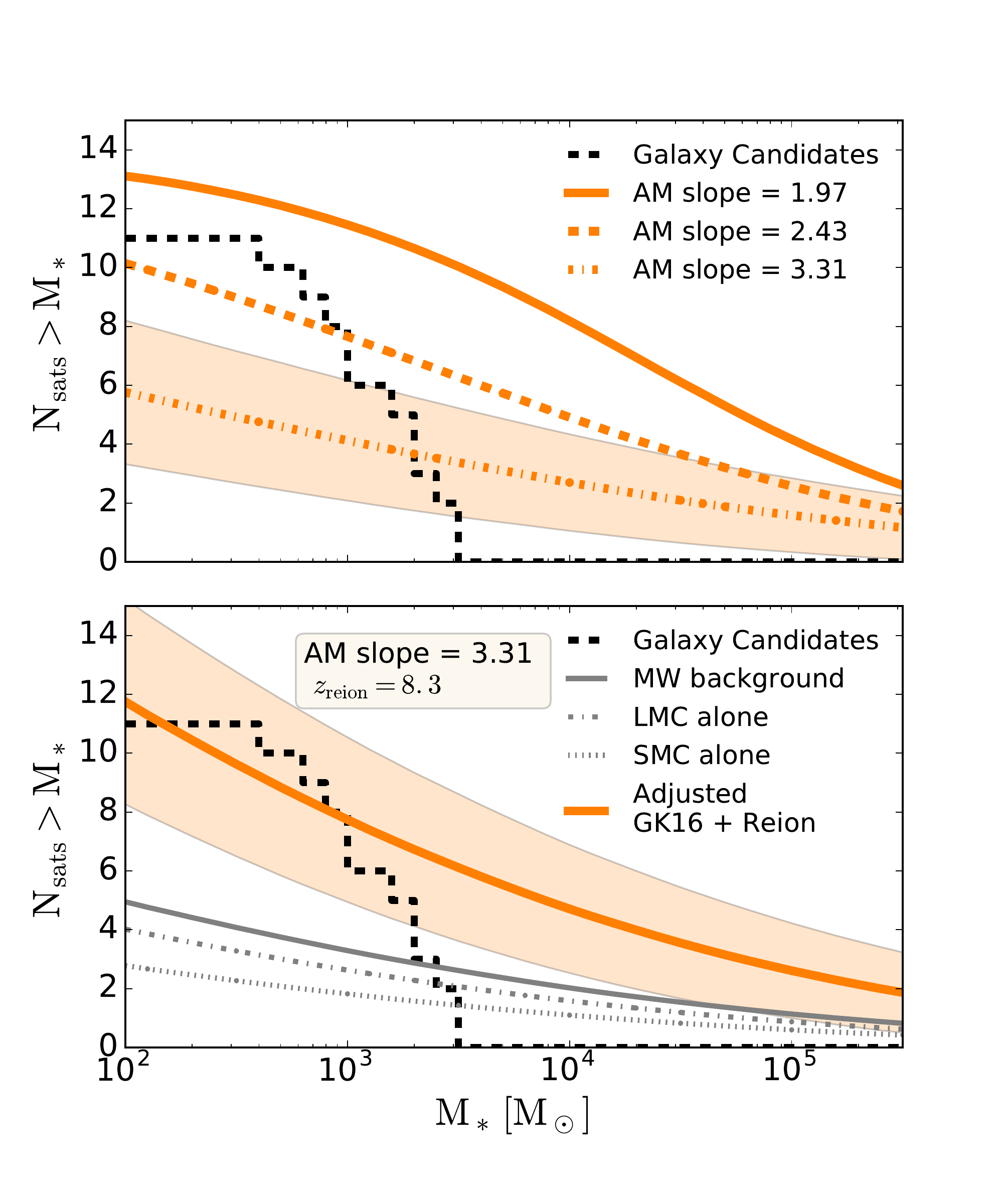}
\caption[Predictions for the LMC vicinity satellite stellar mass function with a steeper \mstarmhalo slope]{\emph{Upper panel: }Predictions for the stellar mass function of satellites within $50$ kpc of the LMC according to the GK16 model with an \mstarmhalo logarithmic slope of $1.97$, $2.43$, and $3.31$. For reference, the Brook model has a slope of $3.1$, but less scatter in the \mstarmhalo relationship. 
One sigma variation is shown around the $3.31$ slope model with a shaded band. \emph{Lower panel: }Same prediction of the GK16 model with a logarithmic slope of $3.31$ except with reionization shifted from $z=13.3$ to $z=8.3$. A similar result could be achieved by lowering the halo size threshold needed for the first galaxies to form.}
\label{fig:steeperslope}
\end{figure}

In the upper panel of Fig.~\ref{fig:steeperslope}, we show our original choice of $\alpha = 1.97$ along with values of $\alpha = 2.43$ and $3.31$ in the GK16 model. The steepest slope greatly diminishes the number of large UFDs, down to $2.6$. As with other possible solutions, however, it also exacerbates the problem of predicting too few small UFDs. 

\vspace{1mm}
\noindent
\newline We elaborate on possible solutions to predicting too few small UFDs in the next three subsections.

\subsubsection*{Reionization occurs later}
Delaying the start of reionization allows for more halos to grow large enough to form stars before reionization suppresses star formation. We illustrate the effect of delaying reionization in Fig.~\ref{fig:tidallystripped}. In the upper panel, the GK16 model with $98\%$ of stellar mass stripped predicts $9$ satellites with $M_* > 10^2$\msun, fewer than the $11$ observed and the $11+$ expected after completion of the MagLiteS and DES surveys. In the lower panel, shifting reionization from $z = 13.3$ to $z=8.3$ increases the number of predicted small UFDs, bringing the prediction into better alignment with observations. Similarly, in the upper panel of Fig.~\ref{fig:steeperslope}, the GK16 model predicts too few small UFDs, especially when the \mstarmhalo relationship slope is $3.31$. In the lower panel, we show again how delaying reionization to $z=8.3$ increases the predicted number of small UFDs. It does, however, also slightly increase the number of large UFDs, making that discrepancy worse.

Recent analysis from the Planck collaboration estimates that reionization occured between $z=7.8$ and $z=8.8$, and that less than $10\%$ of hydrogen in the Universe was ionized before $z=10$ \citep{Planck16b}. This suggests that reionization occuring between $z=15$ and $11.5$ as in \cite{Barber14} is too early, and that shifting reionization to a lower redshift may be part of the solution to predicting more small UFDs.

\subsubsection*{Star formation begins in smaller halos}
Similar to delaying reionization, lowering the size threshold needed for halos to form stars increases the fraction of halos that host luminous galaxies. In our model, this means reducing \vmaxpre. The resultant effect of increasing the number of predicted low mass satellites with $M_* < 10^5$ is nearly identical and therefore degenerate with that of lowering $z_{\rm{reion}}$. Having to lower \vmaxpre would be natural if UFDs formed in $\rm{H_2}$ cooling minihalos before reionization, as hypothesized by \cite{Salvadori09} and \cite{Kirby13}, rather than forming in larger atomic line cooling halos \citep{Bromm11, Power14}.

\subsubsection*{Bent \mstarmhalo relationship}
\begin{figure}
\centering
\includegraphics[width=\figwidth\textwidth]{\figpath 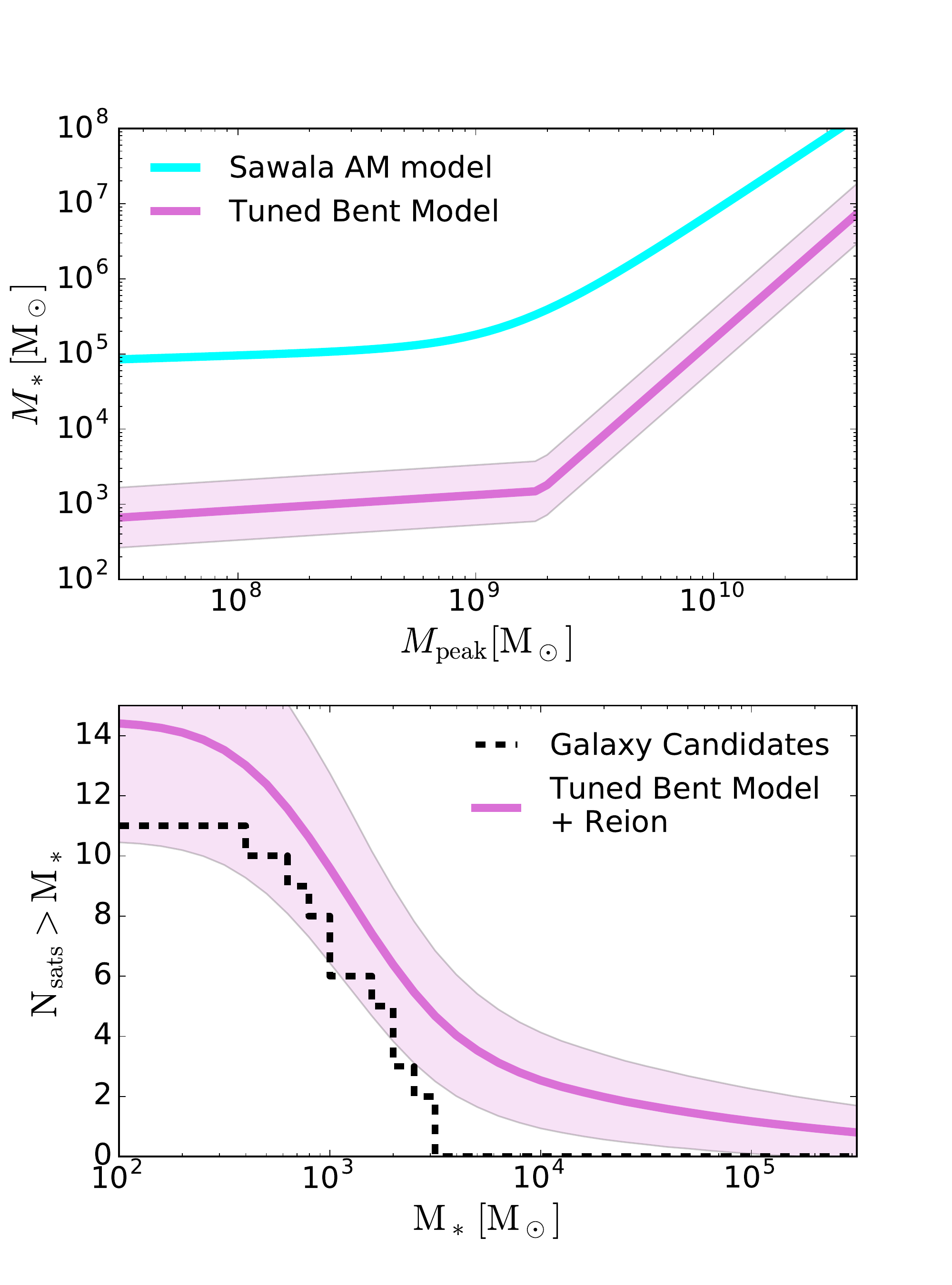}
\caption[Predictions for the LMC vicinity satellite stellar mass function with a bend in \mstarmhalo]{Upper Panel: \mstarmhalo relationships with a bend introduced. The Sawala AM model \protect\citep{Sawala15} derived from hydrodynamic simulations includes a flattening of the relationship for $M_* \lesssim 10^5$\msun. A similar flattening added to the GK16 model with a steep logarithmic slope of $2.77$, except at $M_* \approx 10^3$\msun, is one of multiple ways in which to predict few satellites within $50$ kpc of the LMC that have $M_* > 10^4$\msun, but many with $M_* < 3 \times 10^3$\msun, as is required to best match observed satellites. However, it is also necessary that the scatter in the \mstarmhalo relationship be small in the bend (shown here as $\pm 0.4$ dex), which is inconsistent with that of the GK16 model. Lower Panel: Prediction of the stellar mass function of satellites within $50$ kpc of the LMC according to the bent \mstarmhalo relationship from the upper panel, and our baseline reionization model. A bend with little scatter is one way to create a steep rise in the mass function around $M_* = 10^3$\msun, as is found with observed satellite galaxies.}
\label{fig:bend}
\end{figure}

Increasing the number of small UFDs could also be achieved by introducing a ``bend'' in the \mstarmhalo relationship, as shown in Fig.~\ref{fig:bend}. This would be similar in form to that proposed in \cite{Sawala15} (which uses a mass definition of \mhalo$=$\mpeaktwo), except instead of the \mstarmhalo relationship flattening out near $M_* = 10^5$\msun, it would have to flatten out near $M_* = 10^3$\msun. In fact, the AM model suggested by \cite{Sawala15} would only exacerbate the problem of predicting too many large UFDs. A bend can only increase the number of predicted small UFDs without greatly increasing the number of predicted large UFDs if the scatter in the \mstarmhalo relationship is sufficiently small. In Fig.~\ref{fig:bend}, the scatter is tuned to a constant $\pm 0.4$ dex to best predict the observed stellar mass function (with some room for more discoveries) as shown in the lower panel. If the scatter is as large as that predicted in the baseline GK16 model, $\pm 0.9$ dex at \mpeakvir$= 10^8$\msun, it would once again predict too many large UFDs at a value of $4.1$. A bend at $M_* = 10^3$\msun with little scatter would need to be physically justified. It could, for instance, result from a minimum threshold of stellar mass being created in any single pre-reionization star forming event.

\vspace{4mm}
\noindent
\newline The multitude of ways in which our prediction for the stellar mass function of LMC vicinity satellites can be adjusted leads to innumerable possible solutions when combinations of each effect are considered.  Furthermore, observations are not even complete, and the LMC vicinity represents just a single measurement of a satellite stellar mass function, subject to large variance from counting statistics. We therefore do not propose any single favored model. Instead, we hope identifying the discrepancy and ways to alter predictions will engender future efforts to better explain the issue. One additional valuable constraint that will also improve with future observations is the stellar mass function of UFDs in the full MW volume. Just as adjustments to our models change predictions for the LMC vicinity, they also change predictions for the MW.

\subsubsection{Ratio of MW:LMC:SMC satellites}
\label{sec:ratio}
In spite of shortcomings in predicting the LMC vicinity stellar mass function, the \textit{ratio} of satellites between the MW, LMC and SMC is more certain. In the GK16 model, the ratio of satellites within $50$ kpc of the LMC is $46\%$ from the MW, $33\%$ from the LMC, and $21\%$ from the SMC. The ratio does not change with adjustments to tidal stripping, reionization, nor the slope of the \mstarmhalo relationship. It also does not change with the range of stellar mass considered. In the Brook and Moster models, the ratios are nearly the same at $49\%$ from the MW, $32\%$ from the LMC, and $19\%$ from the SMC. More concretely, of the $12$ UFDs near the LMC, the most consistent ratio of integers would be $6$ or $5$ from the MW, $4$ from the LMC, and $2$ or $3$ from the SMC. This is broadly consistent with \cite{Jethwa16} finding $6/9$ of the UFDs considered within $50$ kpc of the LMC having a $>70\%$ chance of association with the Magellanic Clouds, and \cite{Sales16} finding $6/10$ of them possibly consistent with the Clouds. The fraction of LMC and SMC satellites would decline if their original satellites at infall spread out to larger distances. Changes in the halo mass used for the three galaxies would also adjust the ratio predicted. 

Considering the entire virial volume of the MW, we expect that $15-25\%$ of all MW satellites, regardless of the mass range considered (so long as it is less than that of the SMC), originated within the virial radius of the LMC or SMC before their infall. The Moster and Brook models predict closer to $15\%$, and the GK models closer to $25\%$. From the LMC alone, we expect $10 - 15\%$. This range is higher than the $5\%$ LMC contribution predicted by \cite{Sales16}, who used a lower mass LMC, and within the $1 - 25\%$ range suggested by \cite{Deason15}. The combined Magellanic contribution we predict is less than the $33\%$ predicted by \cite{Jethwa16}.

\subsection{Satellites of isolated LMC analogs}
\label{sec:LMCanalogs}
Isolated LMC-sized galaxies offer a much cleaner way to probe satellite populations. Instead of three overlapping hosts of satellite galaxies in the LMC/SMC/MW case, there is just a single host. Consequently, satellite-host membership is unambiguous, there is no concern about the infall time of the host into another galaxy, and sub-subhalo orbits do not need to be considered. Since isolated LMC-sized galaxies are much less massive than the MW, and satellites can be considered out to the full virial radius beyond $50$~kpc, the average strength of tidal stripping that must be considered is greatly diminished. Also, uncertainty in the radial distribution of galaxies is decreased when considering volumes near the full virial radius. Lastly, measuring many host-satellite population pairs can be used to help control for halo-to-halo variance, and statistical uncertainty in the mass of the hosts. Of the ways suggested to modify the predicted LMC-vicinity satellite stellar mass function presented in Section~\ref{sec:lmcvicinity}, only the \mstarmhalo relationship and reionization remain significant. Therefore, searching for satellites of isolated LMC-sized galaxies can provide valuable information to help constrain models, and to determine how typical the satellite population near our own LMC is.

The majority of known isolated LMC-sized field galaxies are more than $2$ Mpc away, as cataloged in Table~\ref{table:results_lmc}. At that distance, current sensitivities are likely to only find galaxies with $M_* \gtrsim 10^5$\msun. For example, the $M_{\rm V} = -7.7$ ($M_* \sim 10^5$\msun) dwarf at $D\sim3.2$~Mpc discovered by \citet{Carlin16} had only $\sim 25-30$ stars resolved to $\sim 1.5$ mags below the RGB tip \citep[see also][for satellite discoveries around Cen~A at similar detection sensitivities]{Crnojevic16}, which is near the minimum number of resolved stars needed to algorithmically detect galaxies \citep{Walsh09}. Reionization has little impact on galaxies with $M_* > 10^5$\msun, leaving the AM model and Poisson noise as the dominant sources of uncertainty. As such, searching for satellites at current detection thresholds is most helpful in constraining AM models and the slope of the \mstarmhalo relationship. We first focus on satellites with $M_* > 10^5$\msun, then discuss the science return on probing smaller satellites later.

\begin{figure}
\includegraphics[width=\figwidth\textwidth]{\figpath 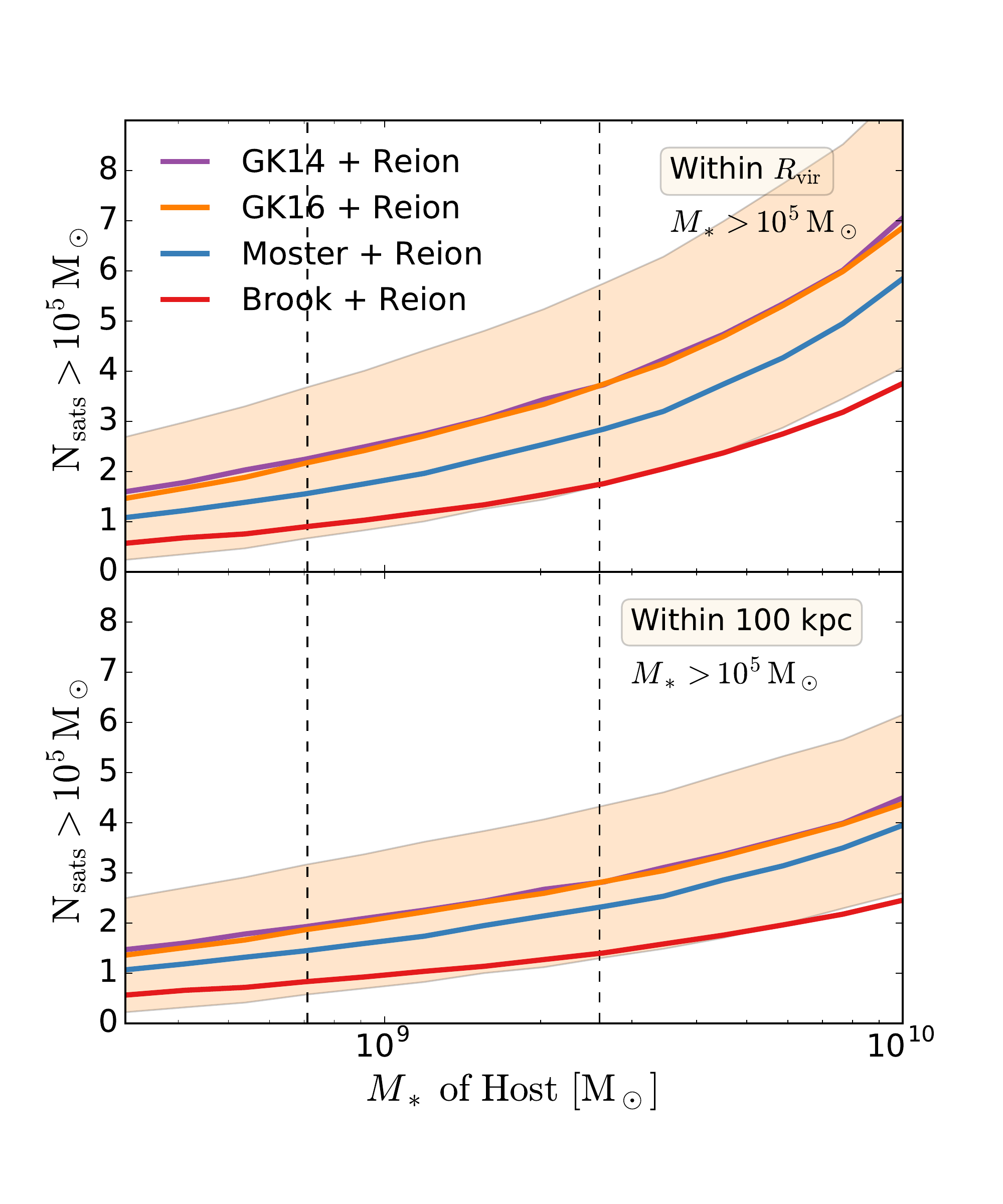}
\caption[Predicted satellite abundance as a function of host stellar mass]{Mean number of satellites with stellar mass above $10^5$\msun as a function of a host halo's total stellar mass, $M_*$, for several AM models. The upper panel shows the number of satellites within the halo's full virial radius. The lower panel shows the number of satellites within $100$ kpc. Within a fixed volume, the number of satellites is less sensitive to the host's total mass. The shape of the curves are the same up to a multiplicative factor for any other stellar mass range considered. $1 \sigma$ halo-to-halo variation is shown for the GK16 model with a shaded band. Vertical lines correspond to the stellar masses of the SMC and the LMC.}
\label{fig:ngreater_lmc}
\end{figure}

As seen in Fig.~\ref{fig:ngreater_minlum_lmc}, within $1\sigma$ about the mean of the GK16 AM model we predict $0.7-3.7$ and $1.7-5.7$ satellites with $M_* > 10^5$\msun around SMC and LMC-sized galaxies respectively. For the Brook model, it drops to $0 - 1.9$ and $0.4-3.1$. If the lack of any satellites with $M_* > 3 \times 10^3$\msun in the LMC vicinity is reflective of typical satellite abundances, the predictions of the Brook model may be favored. Expanding our predictions to isolated hosts over a range of masses, we show in Fig.~\ref{fig:ngreater_lmc} the dependence of satellite abundance on host halo stellar mass. In the upper panel we include all satellites with $M_*>10^5$\msun within the virial volume, while in the lower panel we limit the volume to a sphere of radius $100$ kpc. 

The SHMF is approximately directly proportional to the host halo mass, and $M_{\rm{halo}} \propto M_*^{0.42}$ around both the LMC's and SMC's stellar masses using the GK16 model. Thus doubling the stellar mass of a host results in a halo with $1.34$ times the number of satellites for any satellite mass interval. Much of this increase arises from a larger virial volume being considered. When fixing the volume to $100$ kpc, doubling the halo mass results in a $1.24$ factor of increase in satellites. For the Moster and Brook models, the dependence is slightly steeper with doubling the host stellar mass resulting in $1.4$ ($1.3$) times the satellite abundance within the virial volume ($100$ kpc). Although we only show functions for satellites with $M_* > 10^5$\msun, the shapes of the curves are identical for different stellar mass ranges. Thus the plot shown can be multiplied by a ratio of the abundance of satellites with $M_* > 10^5$\msun to any other range, as can be deduced from Fig.~\ref{fig:ngreater_minlum_lmc}. For instance, the ratio of $M_* > 10^4$\msun to $M_* > 10^5$\msun in the GK16 model is $2.0$ using our baseline reionization model, or $2.6$ using $z_{\rm{reion}} = 9.3$. This apparent sensitivity to reionization makes measuring the ratio of $M_* > 10^5$\msun satellites to lower thresholds a valuable tool for studying the effects of reionization. 

\begin{figure}
\includegraphics[width=\figwidth\textwidth]{\figpath 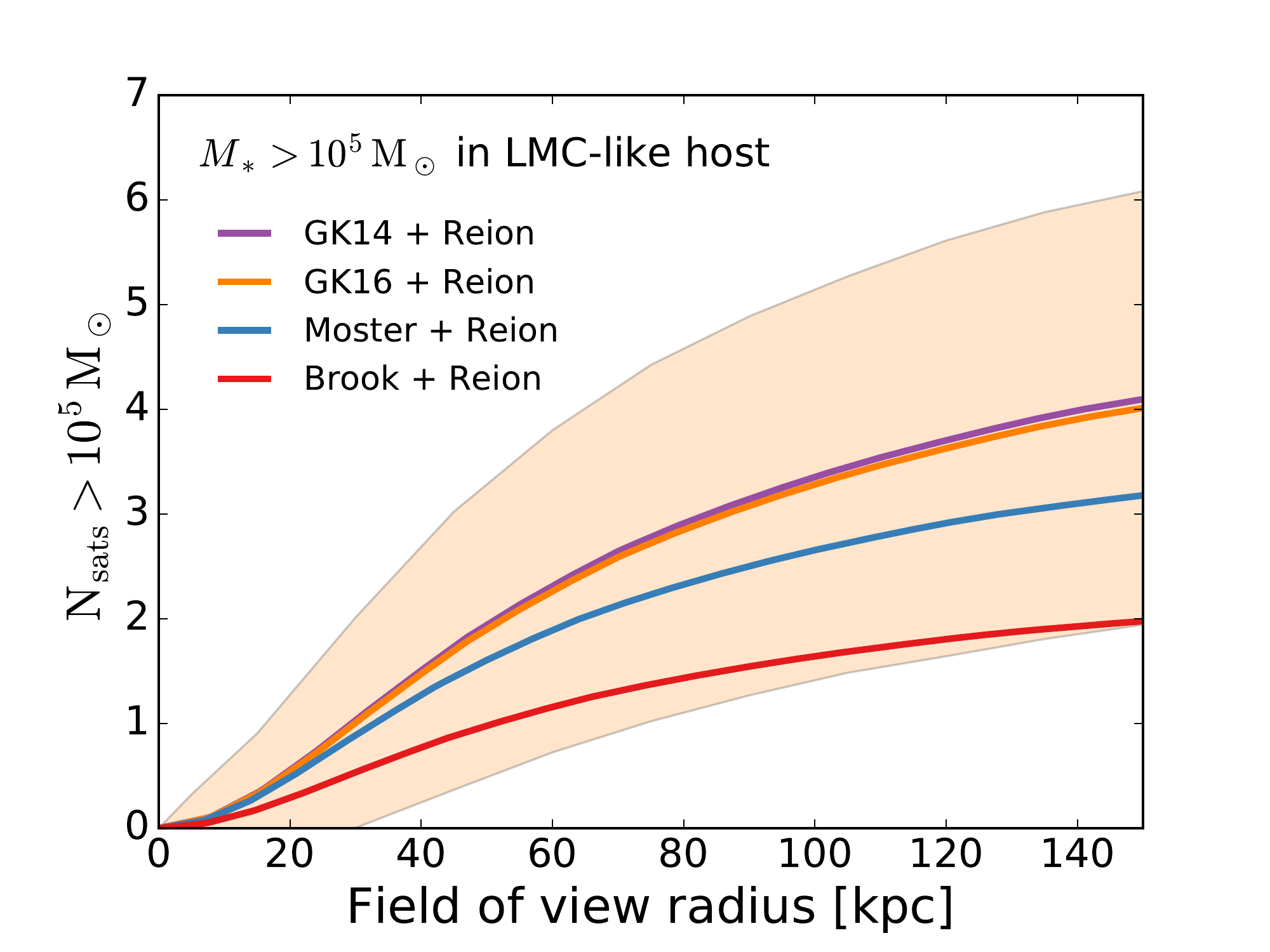}
\caption[Predicted satellite abundance as a function of field of view radius]{Mean number of satellites with $M_* > 10^5$\msun as a function of observed radius perpendicular to the line of sight assuming a circular field of view for an LMC-sized galaxy. We count all halos within $1.5 \times$\rvir of the host in the line of sight direction. $1 \sigma$ halo-to-halo variation is shown for the GK16 model with a shaded band.}
\label{fig:LMC_los_radial}
\end{figure}

When observing distant galaxies, the geometry is dictated by a line of sight, not spherical volumes as considered so far. We give a sense of the cumulative radial distribution of satellites as a function of observed radius perpendicular to the line of sight in physical units for an LMC-sized isolated galaxy in Fig.~\ref{fig:LMC_los_radial}. Using results from \cite{Dooley16b}, we predict how many satellites are within a line of sight with a circular aperture, counting satellites in the axis of the line of sight out to the splashback radius \citep{More15} of $1.5 \times$\rvir. The abundance of satellites is concentrated towards the center, with an observed area out to $0.5 \times$\rvir containing $70\%$ of the total number of satellites within a field of view encompassing the full \rvir. In our LMC-sized galaxy example, observing out to just $50$ kpc is enough to expect $1$ satellite with $M_* > 10^5$\msun in all abundance matching models. Complete information on computing the radial dependence for any host galaxy can be found in \cite{Dooley16b}.

In Table~\ref{table:results_lmc}, we catalog known galaxies with stellar masses on the order of $10^9 - 10^{10}$\msun that are $2-8$ Mpc away, have Galactic latitudes $|b| > 25^\circ$, have low extinction ($E(B-V) < 0.15$), and are categorized as relatively isolated in the Karachentsev et al. catalog. For each galaxy, we indicate the mean number of expected satellites with $M_* > 10^5$\msun within the virial radius. We also list the $20^{th}$ and $80^{th}$ percentile number of satellites, the inferred virial radius of the host, and the mean number of satellites within one pointing of a $1.5^\circ$ and $2.2^\circ$ diameter field of view centered on the host. These angular fields of view correspond to the Hyper SuprimeCam (Subaru 8.2m) and DECam (CTIO Blanco 4m) imagers, the largest current imagers capable of resolving the stellar halos of the galaxies in Table~\ref{table:results_lmc}. We focus on the Brook and GK16 models since the GK14 model is nearly identical to the GK16 model, and the Moster model predictions fall in between those of the Brook and GK16 models. The galaxies are sufficiently distant, and have a large enough predicted satellite population that a single pointing of either camera at each galaxy covers an area expected to contain a mean of at least one observable satellite in most cases. Some of the more distant targets are expected to have as many as four satellites within a single pointing of a $2.2^\circ$ diameter camera in the GK16 model. While a survey of many host galaxies complete to $M_* = 10^5$\msun satellites is needed to test abundance matching models, satellites may be abundant enough that targeting just a few galaxies will reveal new satellites to study in a different environment than the MW.

We note that if the satellite abundance of LMC analogs mimics that of the true LMC, we expect fewer $10^5$\msun galaxies orbiting these systems than predicted by the Brook model. Discovering very few $10^5$\msun satellites in these systems would thus favor a steep \mstarmhalo relationship.

If continued improvement in telescopes, imagers, and survey strategies enables the detection of satellites down to $M_* > 10^4$\msun, we could gain valuable information on how reionization effects galaxy formation, and in turn help unravel the ``small UFD'' problem of the LMC-vicinity puzzle. This is especially true if the \mstarmhalo relationship for $M_* > 10^5$\msun is better constrained, leaving reionization or a bend in the \mstarmhalo relationship as the most significant poorly understood controls on the population of lower mass satellites of isolated LMC-sized hosts. For instance, the ratio of $M_* > 10^4$\msun to $M_* > 10^5$\msun satellites in the GK16 model is $2.0$ using our baseline reionization model, or $2.6$ using $z_{\rm{reion}} = 9.3$. Measuring the ratio of $M_* > 10^5$\msun satellites to lower thresholds is a valuable tool for studying the effects of reionization.

\begin{table*}
\tablewidth{0.97\textwidth}
\centering
\caption{Mean number of observable satellites with $M_* > 10^5$\msun around isolated field dwarf galaxies}
\label{table:results_lmc}
\begin{tabular}{lccccccccccccc} 
\hline
\hline
\textbf{Name} &  \boldmath$M_* $ &  \boldmath$D_{\sun} $ &  \multicolumn{5}{c}{\textbf{Brook}} && \multicolumn{5}{c}{\textbf{GK16}}  \\
 \cline{4-8}\cline{10-14}  \\ 
  & $[10^8 \, \mathrm{M_\odot}]$  &  $[\mathrm{Mpc}]$ & $\bar{N}_{\rm{lum}}$ & 20/80\% & $\bar{N}_{\rm{fov}}^{1.5}$ & $\bar{N}_{\rm{fov}}^{2.2}$ &   $R_{\rm{vir}}$ &   & $\bar{N}_{\rm{lum}}$ & 20/80\% & $\bar{N}_{\rm{fov}}^{1.5}$ & $\bar{N}_{\rm{fov}}^{2.2} $ &   $R_{\rm{vir}}$ \\
\hline
SMC & 7.1 & \nodata     & $0.89$ & $0 / 2$ & \nodata & \nodata & 117 &    & $2.16$ & $1/ 3$ & \nodata & \nodata & 131  \\
\hline
NGC5023 & 10.2 & 6.05     & $1.09$ & $0 / 2$ & $0.97$ & $1.18$ & 124 &    & $2.53$ & $1/ 4$ & $2.11$ & $2.61$ & 138  \\
\hline
NGC5585 & 10.7 & 5.7     & $1.13$ & $0 / 2$ & $0.96$ & $1.18$ & 125 &    & $2.60$ & $1/ 4$ & $2.07$ & $2.58$ & 139  \\
\hline
NGC1800 & 11 & 8     & $1.13$ & $0 / 2$ & $1.16$ & $1.34$ & 126 &    & $2.62$ & $1/ 4$ & $2.54$ & $2.99$ & 139  \\
\hline
ESO383-087 & 11.2 & 3.19     & $1.14$ & $0 / 2$ & $0.60$ & $0.85$ & 126 &    & $2.64$ & $1/ 4$ & $1.26$ & $1.81$ & 140  \\
\hline
NGC5253 & 12 & 3.44     & $1.19$ & $0 / 2$ & $0.67$ & $0.92$ & 127 &    & $2.71$ & $1/ 4$ & $1.39$ & $1.96$ & 141  \\
\hline
NGC4204 & 13.2 & 8     & $1.24$ & $0 / 2$ & $1.25$ & $1.46$ & 129 &    & $2.81$ & $1/ 4$ & $2.69$ & $3.18$ & 143  \\
\hline
HolmII & 15.8 & 3.47     & $1.37$ & $0 / 2$ & $0.75$ & $1.04$ & 134 &    & $3.03$ & $1/ 4$ & $1.51$ & $2.15$ & 146  \\
\hline
NGC4144 & 17.8 & 6.89     & $1.44$ & $0 / 2$ & $1.30$ & $1.57$ & 136 &    & $3.19$ & $2/ 5$ & $2.75$ & $3.37$ & 148  \\
\hline
NGC0404 & 18.2 & 2.98     & $1.46$ & $0 / 2$ & $0.66$ & $0.96$ & 137 &    & $3.20$ & $2/ 5$ & $1.31$ & $1.96$ & 149  \\
\hline
IC5052 & 18.6 & 5.5     & $1.49$ & $0 / 2$ & $1.16$ & $1.46$ & 137 &    & $3.23$ & $2/ 5$ & $2.39$ & $3.04$ & 149  \\
\hline
IC2574 & 21.4 & 3.93     & $1.60$ & $0 / 3$ & $0.94$ & $1.27$ & 140 &    & $3.42$ & $2/ 5$ & $1.85$ & $2.58$ & 152  \\
\hline
NGC0045 & 21.4 & 6.64     & $1.59$ & $0 / 3$ & $1.38$ & $1.69$ & 140 &    & $3.42$ & $2/ 5$ & $2.83$ & $3.50$ & 152  \\
\hline
NGC4136 & 24.6 & 7.9     & $1.71$ & $1 / 3$ & $1.62$ & $1.92$ & 144 &    & $3.64$ & $2/ 5$ & $3.31$ & $3.98$ & 155  \\
\hline
NGC0300 & 25.7 & 2.09     & $1.76$ & $1 / 3$ & $0.47$ & $0.77$ & 145 &    & $3.69$ & $2/ 5$ & $0.88$ & $1.48$ & 156  \\
\hline
LMC & 26 & \nodata     & $1.76$ & $1 / 3$ & \nodata & \nodata & 145 &    & $3.70$ & $2/ 5$ & \nodata & \nodata & 156  \\
\hline
NGC7713 & 26.9 & 7.8     & $1.79$ & $1 / 3$ & $1.67$ & $1.99$ & 146 &    & $3.75$ & $2/ 5$ & $3.37$ & $4.06$ & 157  \\
\hline
NGC4242 & 29.5 & 7.9     & $1.88$ & $1 / 3$ & $1.75$ & $2.09$ & 148 &    & $3.91$ & $2/ 6$ & $3.51$ & $4.24$ & 159  \\
\hline
NGC4395 & 29.5 & 4.76     & $1.89$ & $1 / 3$ & $1.25$ & $1.64$ & 148 &    & $3.89$ & $2/ 6$ & $2.43$ & $3.24$ & 159  \\
\hline
NGC0024 & 30.2 & 7.31     & $1.91$ & $1 / 3$ & $1.70$ & $2.05$ & 149 &    & $3.93$ & $2/ 6$ & $3.36$ & $4.12$ & 159  \\
\hline
NGC0055 & 30.2 & 2.11     & $1.90$ & $1 / 3$ & $0.49$ & $0.81$ & 149 &    & $3.92$ & $2/ 6$ & $0.92$ & $1.55$ & 159  \\
\hline
NGC0247 & 31.6 & 3.72     & $1.96$ & $1 / 3$ & $1.02$ & $1.43$ & 150 &    & $4.02$ & $2/ 6$ & $1.96$ & $2.80$ & 160  \\
\hline
NGC3239 & 33.1 & 7.9     & $2.00$ & $1 / 3$ & $1.84$ & $2.21$ & 151 &    & $4.08$ & $2/ 6$ & $3.62$ & $4.39$ & 161  \\
\hline
NGC4244 & 33.1 & 4.31     & $2.02$ & $1 / 3$ & $1.20$ & $1.62$ & 151 &    & $4.10$ & $2/ 6$ & $2.30$ & $3.17$ & 161  \\
\hline
NGC1313 & 37.1 & 4.31     & $2.13$ & $1 / 3$ & $1.24$ & $1.69$ & 154 &    & $4.30$ & $2/ 6$ & $2.38$ & $3.28$ & 164  \\
\hline
NGC4656 & 38.9 & 5.4     & $2.19$ & $1 / 3$ & $1.55$ & $1.99$ & 156 &    & $4.35$ & $2/ 6$ & $2.93$ & $3.82$ & 165  \\
\hline
NGC4236 & 40.7 & 4.41     & $2.25$ & $1 / 3$ & $1.32$ & $1.80$ & 157 &    & $4.43$ & $3/ 6$ & $2.47$ & $3.41$ & 166  \\
\hline
IC5332 & 41.7 & 7.8     & $2.25$ & $1 / 3$ & $2.01$ & $2.43$ & 158 &    & $4.48$ & $3/ 6$ & $3.88$ & $4.74$ & 166  \\
\hline
NGC4449 & 47.9 & 4.27     & $2.47$ & $1 / 4$ & $1.37$ & $1.89$ & 162 &    & $4.77$ & $3/ 7$ & $2.52$ & $3.53$ & 170  \\
\hline
NGC4605 & 50.1 & 5.55     & $2.50$ & $1 / 4$ & $1.74$ & $2.25$ & 163 &    & $4.87$ & $3/ 7$ & $3.26$ & $4.26$ & 171  \\
\hline
NGC5102 & 50.1 & 3.66     & $2.51$ & $1 / 4$ & $1.18$ & $1.70$ & 163 &    & $4.88$ & $3/ 7$ & $2.17$ & $3.17$ & 171  \\
\hline
NGC7793 & 50.1 & 3.63     & $2.52$ & $1 / 4$ & $1.17$ & $1.69$ & 163 &    & $4.88$ & $3/ 7$ & $2.14$ & $3.15$ & 171  \\
\hline
NGC5068 & 60.3 & 5.45     & $2.79$ & $1 / 4$ & $1.86$ & $2.43$ & 169 &    & $5.30$ & $3/ 7$ & $3.41$ & $4.50$ & 176  \\
\hline
NGC2403 & 72.4 & 3.19     & $3.08$ & $2 / 5$ & $1.13$ & $1.73$ & 174 &    & $5.82$ & $4/ 8$ & $2.03$ & $3.16$ & 181  \\
\hline
\end{tabular}
\vspace{-5mm}
\tablecomments{Mean number of satellites with $M^* > 10^5$\msun expected to exist within the virial volume of known isolated field galaxies as predicted with the Brook and GK16 models. The $20^{\rm{th}}$ and $80^{\rm{th}}$ percentile of the satellite abundance distributions are listed, indicating a $>80\%$ chance of at least one satellite in even the smallest galaxies considered. $\bar{N}_{\rm{fov}}^{1.5}$ and $\bar{N}_{\rm{fov}}^{2.2}$ indicate the mean number of satellites within a field of view of diameter $1.5^{\circ}$ and $2.2^{\circ}$ respectively. These are the apertures of Hyper SuprimeCam and DECam. Also shown is the inferred virial radius of each galaxy.} 
\end{table*}

\section{Conclusions}
\label{sec:conclusions}
Using abundance matching (AM) models, a model for reionization, and simulations from the \textit{Caterpillar} suite, we predict the abundance of satellites around galaxies with stellar masses comparable to the SMC and LMC. We additionally predict the abundance of satellites within $50$ kpc of the actual LMC, modelling the total population as a superposition of satellites originally belonging to the MW, LMC, and SMC. Independent of adjustments to our reionization model, the \mstarmhalo relationship for satellite galaxies, and the stellar mass range of satellites considered, we find the ratio of satellites in the $50$ kpc LMC vicinity to be approximately $47\%$ from the MW, $33\%$ from the LMC, and $20\%$ from the SMC. Moreover, within $300$ kpc of the MW, we estimate $10-15\%$ of all satellites smaller than the SMC were accreted with the LMC, and $5-10\%$ with the SMC. These values are somewhat sensitive to the mass of the host galaxies used, which were \mvir$=1.4\times 10^{12}$\msun for the MW, $\sim 2\times 10^{11}$ for the LMC at infall, and $\sim 1\times 10^{11}$ for the SMC at infall.

Curiously, we note that all twelve currently known satellite candidates within $50$ kpc of the LMC have a stellar mass of $M_* \lesssim 3 \times 10^3$\msun. This creates two significant discrepancies with our predicted satellite stellar mass functions:
\begin{enumerate}
\item
We predict too many $M_* > 10^4$\msun satellites (large UFDs).
\item
We predict too few $10^2 < M_* < 3 \times 10^3$\msun satellites (small UFDs).
\end{enumerate}
For instance, combining our reionization model with the AM model from \cite{GarrisonKimmel16} (GK16), we predict $\sim 8$ large UFDs in the LMC vicinity which has only a $0.04\%$ statistical chance of being consistent with the zero currently known. The same model predicts just $\sim 3$ small UFDs, much less than the $11$ currently known (the twelfth known satellite, Cetus~II, has a stellar mass of just $\sim 80$\msun which is outside our defined range of small UFDs). Furthermore, the problem is only expected to get worse since the full volume under consideration has not yet been surveyed. The ongoing MagLiteS survey is expected to increase the number of known small UFDs near the LMC, but not increase the number of known large UFDs by more than one or two. 

We explore a variety of model adjustments that would produce a prediction more consistent with observations. However, all options have limitations, and combined with incomplete observations, the parameter space of possible adjustments is too large to select any preferred solution. The lack of any known large UFDs near the LMC could for instance be indicative of some combination of a steeper \mstarmhalo relationship, lower mass MW, LMC, and SMC, extreme amount of tidal stripping, a less centrally concentrated radial satellite distribution than we predict, and/or migration of original LMC/SMC satellites to farther distances since infall. The large abundance of small UFDs could be indicative of some combination of reionization occurring later (e.g. $z_{\rm{reion}} \approx 8$ instead of $z_{\rm{reion}} \approx 13$), a reduced \vmax threshold for dark matter halos to first form galaxies (such as UFDs beginning as minihalos instead of atomic cooling halos), or a ``bend'' in the \mstarmhalo relationship at $M_* \approx 10^3$\msun (which could arise from a minimum amount of stellar mass being created in any luminous galaxy). 

We also make predictions for the abundance of satellites of known isolated LMC-sized galaxies which fall between $2$ and $8$ Mpc away. We find that searching these galaxies for satellites is worthwhile not only because of the likelihood of discovering new satellites in different environments, but because it can provide valuable information to help constrain AM models and make better sense of the LMC vicinity satellite population. The stellar mass function of satellites near the LMC cannot be fully understood without a better grasp of the statistical fluctuations in satellite abundances around hosts. 

We find the target galaxies selected likely contain $1-6$ satellites with stellar mass $> 10^5$\msun within their virial volumes. Perhaps more importantly, using a single pointing of a $1.5^{\circ}$ diameter field of view camera is sufficient to expect a mean of $1-3$ such satellites around most targets. If the number of satellites discovered is consistent with or less than predictions from the Brook AM model \citep{Brook14}, as suggested by the stellar mass function of satellites within $50$ kpc of the LMC, it would favor a steep \mstarmhalo relationship. Reionization could not be part of the explanation since it makes little impact on the abundance of satellites with $M_* > 10^5$\msun. Alternatively, if predictions from the GK16 model are more accurate, a steep \mstarmhalo relationship would no longer be a viable part of the explanation for the dearth of large UFDs near the LMC. If surveys are eventually able to discover galaxies down to $M_* = 10^4$\msun or lower, a better understanding of reionization and its ability to suppress star formation can be achieved. If reionization occurs later, or the size threshold needed for galaxies to first form stars is lower, there would be an increase in the ratio of galaxies with $10^4 < M_* < 10^5$\msun to galaxies with $M_* > 10^5$\msun.

The path forward towards understanding the LMC vicinity stellar mass function will require completion of the MagLiteS survey, further analysis of DES data for more galaxies, improved luminosity completeness limits on searches for satellites within the entire MW volume, and conducting searches for satellites around as many other hosts beyond the MW as possible. It will also require improving constraints on reionization and understanding how it influences low mass galaxy formation. Better constraints on the MW mass and orbital histories of the LMC and SMC will additionally help. In spite of many uncertainties, we find that the discrepancies between the predicted and observed LMC-vicinity stellar mass function are surprisingly large and may require significant effort to arrive at a satisfactory solution. We encourage others to join the effort.

\section*{Acknowledgements}

AHGP acknowledges support from National Science Foundation (NSF) Grant No. AST-1615838.   BW acknowledges support by NSF Faculty Early Career Development (CAREER) award AST-1151462. AF acknowledges support from NSF-CAREER grant AST-1255160.


\begin{thebibliography}{}
\makeatletter
\relax
\def\mn@urlcharsother{\let\do\@makeother \do\$\do\&\do\#\do\^\do\_\do\%\do\~}
\def\mn@doi{\begingroup\mn@urlcharsother \@ifnextchar [ {\mn@doi@}
  {\mn@doi@[]}}
\def\mn@doi@[#1]#2{\def\@tempa{#1}\ifx\@tempa\@empty \href
  {http://dx.doi.org/#2} {doi:#2}\else \href {http://dx.doi.org/#2} {#1}\fi
  \endgroup}
\def\mn@eprint#1#2{\mn@eprint@#1:#2::\@nil}
\def\mn@eprint@arXiv#1{\href {http://arxiv.org/abs/#1} {{\tt arXiv:#1}}}
\def\mn@eprint@dblp#1{\href {http://dblp.uni-trier.de/rec/bibtex/#1.xml}
  {dblp:#1}}
\def\mn@eprint@#1:#2:#3:#4\@nil{\def\@tempa {#1}\def\@tempb {#2}\def\@tempc
  {#3}\ifx \@tempc \@empty \let \@tempc \@tempb \let \@tempb \@tempa \fi \ifx
  \@tempb \@empty \def\@tempb {arXiv}\fi \@ifundefined
  {mn@eprint@\@tempb}{\@tempb:\@tempc}{\expandafter \expandafter \csname
  mn@eprint@\@tempb\endcsname \expandafter{\@tempc}}}

\bibitem[\protect\citeauthoryear{{Barber}, {Starkenburg}, {Navarro},
  {McConnachie}  \& {Fattahi}}{{Barber} et~al.}{2014}]{Barber14}
{Barber} C.,  {Starkenburg} E.,  {Navarro} J.~F.,  {McConnachie} A.~W.,
  {Fattahi} A.,  2014, \mn@doi [\mnras] {10.1093/mnras/stt1959}, \href
  {http://adsabs.harvard.edu/abs/2014MNRAS.437..959B} {437, 959}

\bibitem[\protect\citeauthoryear{{Bechtol} et~al.,}{{Bechtol}
  et~al.}{2015}]{Bechtol15}
{Bechtol} K.,  et~al., 2015, \mn@doi [\apj] {10.1088/0004-637X/807/1/50}, \href
  {http://adsabs.harvard.edu/abs/2015ApJ...807...50B} {807, 50}

\bibitem[\protect\citeauthoryear{{Behroozi}, {Wechsler}  \&
  {Conroy}}{{Behroozi} et~al.}{2013}]{Behroozi13}
{Behroozi} P.~S.,  {Wechsler} R.~H.,   {Conroy} C.,  2013, \mn@doi [\apj]
  {10.1088/0004-637X/770/1/57}, \href
  {http://adsabs.harvard.edu/abs/2013ApJ...770...57B} {770, 57}

\bibitem[\protect\citeauthoryear{{Bekki}}{{Bekki}}{2011}]{Bekki11}
{Bekki} K.,  2011, \mn@doi [\mnras] {10.1111/j.1365-2966.2011.19211.x}, \href
  {http://adsabs.harvard.edu/abs/2011MNRAS.416.2359B} {416, 2359}

\bibitem[\protect\citeauthoryear{{Bellazzini}, {Ibata}, {Monaco}, {Martin},
  {Irwin}  \& {Lewis}}{{Bellazzini} et~al.}{2004}]{Bellazzini04}
{Bellazzini} M.,  {Ibata} R.,  {Monaco} L.,  {Martin} N.,  {Irwin} M.~J.,
  {Lewis} G.~F.,  2004, \mn@doi [\mnras] {10.1111/j.1365-2966.2004.08283.x},
  \href {http://adsabs.harvard.edu/abs/2004MNRAS.354.1263B} {354, 1263}

\bibitem[\protect\citeauthoryear{{Bellazzini}, {Ibata}, {Martin}, {Lewis},
  {Conn}  \& {Irwin}}{{Bellazzini} et~al.}{2006}]{Bellazzini06}
{Bellazzini} M.,  {Ibata} R.,  {Martin} N.,  {Lewis} G.~F.,  {Conn} B.,
  {Irwin} M.~J.,  2006, \mn@doi [\mnras] {10.1111/j.1365-2966.2005.09973.x},
  \href {http://adsabs.harvard.edu/abs/2006MNRAS.366..865B} {366, 865}

\bibitem[\protect\citeauthoryear{{Besla}, {Kallivayalil}, {Hernquist},
  {Robertson}, {Cox}, {van der Marel}  \& {Alcock}}{{Besla}
  et~al.}{2007}]{Besla07}
{Besla} G.,  {Kallivayalil} N.,  {Hernquist} L.,  {Robertson} B.,  {Cox} T.~J.,
   {van der Marel} R.~P.,   {Alcock} C.,  2007, \mn@doi [\apj]
  {10.1086/521385}, \href {http://adsabs.harvard.edu/abs/2007ApJ...668..949B}
  {668, 949}

\bibitem[\protect\citeauthoryear{{Bovill} \& {Ricotti}}{{Bovill} \&
  {Ricotti}}{2009}]{Bovill09}
{Bovill} M.~S.,  {Ricotti} M.,  2009, \mn@doi [\apj]
  {10.1088/0004-637X/693/2/1859}, \href
  {http://adsabs.harvard.edu/abs/2009ApJ...693.1859B} {693, 1859}

\bibitem[\protect\citeauthoryear{{Boylan-Kolchin}, {Springel}, {White}  \&
  {Jenkins}}{{Boylan-Kolchin} et~al.}{2010}]{BoylanKolchin10}
{Boylan-Kolchin} M.,  {Springel} V.,  {White} S.~D.~M.,   {Jenkins} A.,  2010,
  \mn@doi [\mnras] {10.1111/j.1365-2966.2010.16774.x}, \href
  {http://adsabs.harvard.edu/abs/2010MNRAS.406..896B} {406, 896}

\bibitem[\protect\citeauthoryear{{Bromm} \& {Yoshida}}{{Bromm} \&
  {Yoshida}}{2011}]{Bromm11}
{Bromm} V.,  {Yoshida} N.,  2011, \mn@doi [\araa]
  {10.1146/annurev-astro-081710-102608}, \href
  {http://adsabs.harvard.edu/abs/2011ARA%26A..49..373B} {49, 373}

\bibitem[\protect\citeauthoryear{{Brook}, {Di Cintio}, {Knebe},
  {Gottl{\"o}ber}, {Hoffman}, {Yepes}  \& {Garrison-Kimmel}}{{Brook}
  et~al.}{2014}]{Brook14}
{Brook} C.~B.,  {Di Cintio} A.,  {Knebe} A.,  {Gottl{\"o}ber} S.,  {Hoffman}
  Y.,  {Yepes} G.,   {Garrison-Kimmel} S.,  2014, \mn@doi [\apjl]
  {10.1088/2041-8205/784/1/L14}, \href
  {http://adsabs.harvard.edu/abs/2014ApJ...784L..14B} {784, L14}

\bibitem[\protect\citeauthoryear{{Brooks}, {Kuhlen}, {Zolotov}  \&
  {Hooper}}{{Brooks} et~al.}{2013}]{Brooks13}
{Brooks} A.~M.,  {Kuhlen} M.,  {Zolotov} A.,   {Hooper} D.,  2013, \mn@doi
  [\apj] {10.1088/0004-637X/765/1/22}, \href
  {http://adsabs.harvard.edu/abs/2013ApJ...765...22B} {765, 22}

\bibitem[\protect\citeauthoryear{{Brown} et~al.,}{{Brown}
  et~al.}{2012}]{Brown12}
{Brown} T.~M.,  et~al., 2012, \mn@doi [\apjl] {10.1088/2041-8205/753/1/L21},
  \href {http://adsabs.harvard.edu/abs/2012ApJ...753L..21B} {753, L21}

\bibitem[\protect\citeauthoryear{{Brown} et~al.,}{{Brown}
  et~al.}{2014a}]{Brown14a}
{Brown} T.~M.,  et~al., 2014a, \memsai, \href
  {http://adsabs.harvard.edu/abs/2014MmSAI..85..493B} {85, 493}

\bibitem[\protect\citeauthoryear{{Brown} et~al.,}{{Brown}
  et~al.}{2014b}]{Brown14b}
{Brown} T.~M.,  et~al., 2014b, \mn@doi [\apj] {10.1088/0004-637X/796/2/91},
  \href {http://adsabs.harvard.edu/abs/2014ApJ...796...91B} {796, 91}

\bibitem[\protect\citeauthoryear{{Bryan} \& {Norman}}{{Bryan} \&
  {Norman}}{1998}]{Bryan98}
{Bryan} G.~L.,  {Norman} M.~L.,  1998, \mn@doi [\apj] {10.1086/305262}, \href
  {http://adsabs.harvard.edu/abs/1998ApJ...495...80B} {495, 80}

\bibitem[\protect\citeauthoryear{{Busha}, {Marshall}, {Wechsler}, {Klypin}  \&
  {Primack}}{{Busha} et~al.}{2011}]{Busha11b}
{Busha} M.~T.,  {Marshall} P.~J.,  {Wechsler} R.~H.,  {Klypin} A.,   {Primack}
  J.,  2011, \mn@doi [\apj] {10.1088/0004-637X/743/1/40}, \href
  {http://adsabs.harvard.edu/abs/2011ApJ...743...40B} {743, 40}

\bibitem[\protect\citeauthoryear{{Carlin} et~al.,}{{Carlin}
  et~al.}{2016}]{Carlin16}
{Carlin} J.~L.,  et~al., 2016, \mn@doi [\apjl] {10.3847/2041-8205/828/1/L5},
  \href {http://adsabs.harvard.edu/abs/2016ApJ...828L...5C} {828, L5}

\bibitem[\protect\citeauthoryear{{Chiboucas}, {Jacobs}, {Tully}  \&
  {Karachentsev}}{{Chiboucas} et~al.}{2013}]{Chiboucas13}
{Chiboucas} K.,  {Jacobs} B.~A.,  {Tully} R.~B.,   {Karachentsev} I.~D.,  2013,
  \mn@doi [\aj] {10.1088/0004-6256/146/5/126}, \href
  {http://adsabs.harvard.edu/abs/2013AJ....146..126C} {146, 126}

\bibitem[\protect\citeauthoryear{{Collins}, {Tollerud}, {Sand}, {Bonaca},
  {Willman}  \& {Strader}}{{Collins} et~al.}{2016}]{Collins16}
{Collins} M.~L.~M.,  {Tollerud} E.~J.,  {Sand} D.~J.,  {Bonaca} A.,  {Willman}
  B.,   {Strader} J.,  2016, preprint, \href
  {http://adsabs.harvard.edu/abs/2016arXiv160805710C} {} (\mn@eprint {arXiv}
  {1608.05710})

\bibitem[\protect\citeauthoryear{{Crnojevi{\'c}} et~al.,}{{Crnojevi{\'c}}
  et~al.}{2014}]{Crnojevic14}
{Crnojevi{\'c}} D.,  et~al., 2014, \mn@doi [\apjl]
  {10.1088/2041-8205/795/2/L35}, \href
  {http://adsabs.harvard.edu/abs/2014ApJ...795L..35C} {795, L35}

\bibitem[\protect\citeauthoryear{{Crnojevi{\'c}} et~al.,}{{Crnojevi{\'c}}
  et~al.}{2016}]{Crnojevic16}
{Crnojevi{\'c}} D.,  et~al., 2016, \mn@doi [\apj] {10.3847/0004-637X/823/1/19},
  \href {http://adsabs.harvard.edu/abs/2016ApJ...823...19C} {823, 19}

\bibitem[\protect\citeauthoryear{{D'Onghia} \& {Lake}}{{D'Onghia} \&
  {Lake}}{2008}]{DOnghia08}
{D'Onghia} E.,  {Lake} G.,  2008, \mn@doi [\apjl] {10.1086/592995}, \href
  {http://adsabs.harvard.edu/abs/2008ApJ...686L..61D} {686, L61}

\bibitem[\protect\citeauthoryear{{Deason}, {Wetzel}, {Garrison-Kimmel}  \&
  {Belokurov}}{{Deason} et~al.}{2015}]{Deason15}
{Deason} A.~J.,  {Wetzel} A.~R.,  {Garrison-Kimmel} S.,   {Belokurov} V.,
  2015, \mn@doi [\mnras] {10.1093/mnras/stv1939}, \href
  {http://adsabs.harvard.edu/abs/2015MNRAS.453.3568D} {453, 3568}

\bibitem[\protect\citeauthoryear{{Dooley}, {Peter}, {Yang}, {Willman},
  {Griffen}  \& {Frebel}}{{Dooley} et~al.}{2016a}]{Dooley16b}
{Dooley} G.~A.,  {Peter} A.~H.~G.,  {Yang} T.,  {Willman} B.,  {Griffen} B.~F.,
    {Frebel} A.,  2016a, preprint, \href
  {http://adsabs.harvard.edu/abs/2016arXiv161000708D} {} (\mn@eprint {arXiv}
  {1610.00708})

\bibitem[\protect\citeauthoryear{{Dooley}, {Peter}, {Vogelsberger}, {Zavala}
  \& {Frebel}}{{Dooley} et~al.}{2016b}]{Dooley16}
{Dooley} G.~A.,  {Peter} A.~H.~G.,  {Vogelsberger} M.,  {Zavala} J.,   {Frebel}
  A.,  2016b, \mn@doi [\mnras] {10.1093/mnras/stw1309}, \href
  {http://adsabs.harvard.edu/abs/2016MNRAS.461..710D} {461, 710}

\bibitem[\protect\citeauthoryear{{Drlica-Wagner} et~al.,}{{Drlica-Wagner}
  et~al.}{2015}]{Drlica15}
{Drlica-Wagner} A.,  et~al., 2015, \mn@doi [\apj]
  {10.1088/0004-637X/813/2/109}, \href
  {http://adsabs.harvard.edu/abs/2015ApJ...813..109D} {813, 109}

\bibitem[\protect\citeauthoryear{{Drlica-Wagner} et~al.,}{{Drlica-Wagner}
  et~al.}{2016}]{Drlica-Wagner16}
{Drlica-Wagner} A.,  et~al., 2016, preprint, \href
  {http://adsabs.harvard.edu/abs/2016arXiv160902148D} {} (\mn@eprint {arXiv}
  {1609.02148})

\bibitem[\protect\citeauthoryear{{Fermi-LAT} et~al.,}{{Fermi-LAT}
  et~al.}{2016}]{FermiLAT16}
{Fermi-LAT} T.,  et~al., 2016, preprint, \href
  {http://adsabs.harvard.edu/abs/2016arXiv161103184F} {} (\mn@eprint {arXiv}
  {1611.03184})

\bibitem[\protect\citeauthoryear{{Garrison-Kimmel}, {Boylan-Kolchin}, {Bullock}
   \& {Lee}}{{Garrison-Kimmel} et~al.}{2014}]{GarrisonKimmel14}
{Garrison-Kimmel} S.,  {Boylan-Kolchin} M.,  {Bullock} J.~S.,   {Lee} K.,
  2014, \mn@doi [\mnras] {10.1093/mnras/stt2377}, \href
  {http://adsabs.harvard.edu/abs/2014MNRAS.438.2578G} {438, 2578}

\bibitem[\protect\citeauthoryear{{Garrison-Kimmel}, {Bullock}, {Boylan-Kolchin}
   \& {Bardwell}}{{Garrison-Kimmel} et~al.}{2016}]{GarrisonKimmel16}
{Garrison-Kimmel} S.,  {Bullock} J.~S.,  {Boylan-Kolchin} M.,   {Bardwell} E.,
  2016, preprint, \href {http://adsabs.harvard.edu/abs/2016arXiv160304855G} {}
  (\mn@eprint {arXiv} {1603.04855})

\bibitem[\protect\citeauthoryear{{Garrison-Kimmel} et~al.,}{{Garrison-Kimmel}
  et~al.}{2017}]{GarrisonKimmel17}
{Garrison-Kimmel} S.,  et~al., 2017, \mn@doi [\mnras] {10.1093/mnras/stx1710},
  \href {http://adsabs.harvard.edu/abs/2017MNRAS.471.1709G} {471, 1709}

\bibitem[\protect\citeauthoryear{{Governato} et~al.,}{{Governato}
  et~al.}{2010}]{Governato10}
{Governato} F.,  et~al., 2010, \mn@doi [\nat] {10.1038/nature08640}, \href
  {http://adsabs.harvard.edu/abs/2010Natur.463..203G} {463, 203}

\bibitem[\protect\citeauthoryear{{Griffen}, {Ji}, {Dooley}, {G{\'o}mez},
  {Vogelsberger}, {O'Shea}  \& {Frebel}}{{Griffen} et~al.}{2016}]{Griffen16}
{Griffen} B.~F.,  {Ji} A.~P.,  {Dooley} G.~A.,  {G{\'o}mez} F.~A.,
  {Vogelsberger} M.,  {O'Shea} B.~W.,   {Frebel} A.,  2016, \mn@doi [\apj]
  {10.3847/0004-637X/818/1/10}, \href
  {http://adsabs.harvard.edu/abs/2016ApJ...818...10G} {818, 10}

\bibitem[\protect\citeauthoryear{{Hargis}, {Willman}  \& {Peter}}{{Hargis}
  et~al.}{2014}]{Hargis14}
{Hargis} J.~R.,  {Willman} B.,   {Peter} A.~H.~G.,  2014, \mn@doi [\apjl]
  {10.1088/2041-8205/795/1/L13}, \href
  {http://adsabs.harvard.edu/abs/2014ApJ...795L..13H} {795, L13}

\bibitem[\protect\citeauthoryear{{Harris} \& {Zaritsky}}{{Harris} \&
  {Zaritsky}}{2009}]{Harris09}
{Harris} J.,  {Zaritsky} D.,  2009, \mn@doi [\aj]
  {10.1088/0004-6256/138/5/1243}, \href
  {http://adsabs.harvard.edu/abs/2009AJ....138.1243H} {138, 1243}

\bibitem[\protect\citeauthoryear{{Hernquist}}{{Hernquist}}{1990}]{Hernquist90}
{Hernquist} L.,  1990, \mn@doi [\apj] {10.1086/168845}, \href
  {http://adsabs.harvard.edu/abs/1990ApJ...356..359H} {356, 359}

\bibitem[\protect\citeauthoryear{{Homma} et~al.,}{{Homma}
  et~al.}{2016}]{Homma16}
{Homma} D.,  et~al., 2016, preprint, \href
  {http://adsabs.harvard.edu/abs/2016arXiv160904346H} {} (\mn@eprint {arXiv}
  {1609.04346})

\bibitem[\protect\citeauthoryear{{Jethwa}, {Belokurov}  \& {Erkal}}{{Jethwa}
  et~al.}{2016a}]{Jethwa16b}
{Jethwa} P.,  {Belokurov} V.,   {Erkal} D.,  2016a, preprint, \href
  {http://adsabs.harvard.edu/abs/2016arXiv161207834J} {} (\mn@eprint {arXiv}
  {1612.07834})

\bibitem[\protect\citeauthoryear{{Jethwa}, {Erkal}  \& {Belokurov}}{{Jethwa}
  et~al.}{2016b}]{Jethwa16}
{Jethwa} P.,  {Erkal} D.,   {Belokurov} V.,  2016b, \mn@doi [\mnras]
  {10.1093/mnras/stw1343}, \href
  {http://adsabs.harvard.edu/abs/2016MNRAS.461.2212J} {461, 2212}

\bibitem[\protect\citeauthoryear{{Kallivayalil}, {van der Marel}, {Besla},
  {Anderson}  \& {Alcock}}{{Kallivayalil} et~al.}{2013}]{Kallivayalil13}
{Kallivayalil} N.,  {van der Marel} R.~P.,  {Besla} G.,  {Anderson} J.,
  {Alcock} C.,  2013, \mn@doi [\apj] {10.1088/0004-637X/764/2/161}, \href
  {http://adsabs.harvard.edu/abs/2013ApJ...764..161K} {764, 161}

\bibitem[\protect\citeauthoryear{{Karachentsev}, {Makarov}  \&
  {Kaisina}}{{Karachentsev} et~al.}{2013}]{Karachentsev13}
{Karachentsev} I.~D.,  {Makarov} D.~I.,   {Kaisina} E.~I.,  2013, \mn@doi [\aj]
  {10.1088/0004-6256/145/4/101}, \href
  {http://adsabs.harvard.edu/abs/2013AJ....145..101K} {145, 101}

\bibitem[\protect\citeauthoryear{{Kim} \& {Jerjen}}{{Kim} \&
  {Jerjen}}{2015}]{Kim15b}
{Kim} D.,  {Jerjen} H.,  2015, \mn@doi [\apjl] {10.1088/2041-8205/808/2/L39},
  \href {http://adsabs.harvard.edu/abs/2015ApJ...808L..39K} {808, L39}

\bibitem[\protect\citeauthoryear{{Kim}, {Jerjen}, {Mackey}, {Da Costa}  \&
  {Milone}}{{Kim} et~al.}{2015}]{Kim15a}
{Kim} D.,  {Jerjen} H.,  {Mackey} D.,  {Da Costa} G.~S.,   {Milone} A.~P.,
  2015, \mn@doi [\apjl] {10.1088/2041-8205/804/2/L44}, \href
  {http://adsabs.harvard.edu/abs/2015ApJ...804L..44K} {804, L44}

\bibitem[\protect\citeauthoryear{{Kirby}, {Boylan-Kolchin}, {Cohen}, {Geha},
  {Bullock}  \& {Kaplinghat}}{{Kirby} et~al.}{2013a}]{Kirby13}
{Kirby} E.~N.,  {Boylan-Kolchin} M.,  {Cohen} J.~G.,  {Geha} M.,  {Bullock}
  J.~S.,   {Kaplinghat} M.,  2013a, \mn@doi [\apj]
  {10.1088/0004-637X/770/1/16}, \href
  {http://adsabs.harvard.edu/abs/2013ApJ...770...16K} {770, 16}

\bibitem[\protect\citeauthoryear{{Kirby}, {Cohen}, {Guhathakurta}, {Cheng},
  {Bullock}  \& {Gallazzi}}{{Kirby} et~al.}{2013b}]{Kirby13b}
{Kirby} E.~N.,  {Cohen} J.~G.,  {Guhathakurta} P.,  {Cheng} L.,  {Bullock}
  J.~S.,   {Gallazzi} A.,  2013b, \mn@doi [\apj] {10.1088/0004-637X/779/2/102},
  \href {http://adsabs.harvard.edu/abs/2013ApJ...779..102K} {779, 102}

\bibitem[\protect\citeauthoryear{{Koposov}, {Belokurov}, {Torrealba}  \&
  {Evans}}{{Koposov} et~al.}{2015a}]{Koposov15}
{Koposov} S.~E.,  {Belokurov} V.,  {Torrealba} G.,   {Evans} N.~W.,  2015a,
  \mn@doi [\apj] {10.1088/0004-637X/805/2/130}, \href
  {http://adsabs.harvard.edu/abs/2015ApJ...805..130K} {805, 130}

\bibitem[\protect\citeauthoryear{{Koposov} et~al.,}{{Koposov}
  et~al.}{2015b}]{Koposov15b}
{Koposov} S.~E.,  et~al., 2015b, \mn@doi [\apj] {10.1088/0004-637X/811/1/62},
  \href {http://adsabs.harvard.edu/abs/2015ApJ...811...62K} {811, 62}

\bibitem[\protect\citeauthoryear{{K{\"u}pper}, {Johnston}, {Mieske}, {Collins}
  \& {Tollerud}}{{K{\"u}pper} et~al.}{2016}]{Kupper16}
{K{\"u}pper} A.~H.~W.,  {Johnston} K.~V.,  {Mieske} S.,  {Collins} M.~L.~M.,
  {Tollerud} E.~J.,  2016, preprint, \href
  {http://adsabs.harvard.edu/abs/2016arXiv160805085K} {} (\mn@eprint {arXiv}
  {1608.05085})

\bibitem[\protect\citeauthoryear{{Laevens} et~al.,}{{Laevens}
  et~al.}{2015}]{Laevens15}
{Laevens} B.~P.~M.,  et~al., 2015, \mn@doi [\apj] {10.1088/0004-637X/813/1/44},
  \href {http://adsabs.harvard.edu/abs/2015ApJ...813...44L} {813, 44}

\bibitem[\protect\citeauthoryear{{L{\'o}pez-Corredoira}}{{L{\'o}pez-Corredoira}}{2006}]{LopezCorredoira06}
{L{\'o}pez-Corredoira} M.,  2006, \mn@doi [\mnras]
  {10.1111/j.1365-2966.2006.10435.x}, \href
  {http://adsabs.harvard.edu/abs/2006MNRAS.369.1911L} {369, 1911}

\bibitem[\protect\citeauthoryear{{Lu}, {Benson}, {Mao}, {Tonnesen}, {Peter},
  {Wetzel}, {Boylan-Kolchin}  \& {Wechsler}}{{Lu} et~al.}{2016}]{Lu16}
{Lu} Y.,  {Benson} A.,  {Mao} Y.-Y.,  {Tonnesen} S.,  {Peter} A.~H.~G.,
  {Wetzel} A.~R.,  {Boylan-Kolchin} M.,   {Wechsler} R.~H.,  2016, preprint,
  \href {http://adsabs.harvard.edu/abs/2016arXiv160502075L} {} (\mn@eprint
  {arXiv} {1605.02075})

\bibitem[\protect\citeauthoryear{{Luque} et~al.,}{{Luque}
  et~al.}{2016}]{Luque16}
{Luque} E.,  et~al., 2016, preprint, \href
  {http://adsabs.harvard.edu/abs/2016arXiv160804033L} {} (\mn@eprint {arXiv}
  {1608.04033})

\bibitem[\protect\citeauthoryear{{Lynden-Bell}}{{Lynden-Bell}}{1976}]{Lynden-Bell76}
{Lynden-Bell} D.,  1976, in {Dickens} R.~J.,  {Perry} J.~E.,  {Smith} F.~G.,
  {King} I.~R.,  eds,  Royal Greenwich Observatory Bulletins Vol. 182, The
  Galaxy and the Local Group. p.~235

\bibitem[\protect\citeauthoryear{{Lynden-Bell}}{{Lynden-Bell}}{1982}]{Lynden-Bell82}
{Lynden-Bell} D.,  1982, The Observatory, \href
  {http://adsabs.harvard.edu/abs/1982Obs...102....7L} {102, 7}

\bibitem[\protect\citeauthoryear{{Lynden-Bell} \& {Lynden-Bell}}{{Lynden-Bell}
  \& {Lynden-Bell}}{1995}]{Lynden-Bell95}
{Lynden-Bell} D.,  {Lynden-Bell} R.~M.,  1995, \mn@doi [\mnras]
  {10.1093/mnras/275.2.429}, \href
  {http://adsabs.harvard.edu/abs/1995MNRAS.275..429L} {275, 429}

\bibitem[\protect\citeauthoryear{{Mao}, {Williamson}  \& {Wechsler}}{{Mao}
  et~al.}{2015}]{Mao15}
{Mao} Y.-Y.,  {Williamson} M.,   {Wechsler} R.~H.,  2015, \mn@doi [\apj]
  {10.1088/0004-637X/810/1/21}, \href
  {http://adsabs.harvard.edu/abs/2015ApJ...810...21M} {810, 21}

\bibitem[\protect\citeauthoryear{{Martin}, {Ibata}, {Bellazzini}, {Irwin},
  {Lewis}  \& {Dehnen}}{{Martin} et~al.}{2004}]{Martin04}
{Martin} N.~F.,  {Ibata} R.~A.,  {Bellazzini} M.,  {Irwin} M.~J.,  {Lewis}
  G.~F.,   {Dehnen} W.,  2004, \mn@doi [\mnras]
  {10.1111/j.1365-2966.2004.07331.x}, \href
  {http://adsabs.harvard.edu/abs/2004MNRAS.348...12M} {348, 12}

\bibitem[\protect\citeauthoryear{{Martin} et~al.,}{{Martin}
  et~al.}{2015}]{Martin15}
{Martin} N.~F.,  et~al., 2015, \mn@doi [\apjl] {10.1088/2041-8205/804/1/L5},
  \href {http://adsabs.harvard.edu/abs/2015ApJ...804L...5M} {804, L5}

\bibitem[\protect\citeauthoryear{{Mart{\'{\i}}nez-Delgado}, {Butler}, {Rix},
  {Franco}, {Pe{\~n}arrubia}, {Alfaro}  \& {Dinescu}}{{Mart{\'{\i}}nez-Delgado}
  et~al.}{2005}]{MartinezDelgado05}
{Mart{\'{\i}}nez-Delgado} D.,  {Butler} D.~J.,  {Rix} H.-W.,  {Franco} V.~I.,
  {Pe{\~n}arrubia} J.,  {Alfaro} E.~J.,   {Dinescu} D.~I.,  2005, \mn@doi
  [\apj] {10.1086/432635}, \href
  {http://adsabs.harvard.edu/abs/2005ApJ...633..205M} {633, 205}

\bibitem[\protect\citeauthoryear{{Mart{\'{\i}}nez-Delgado}
  et~al.,}{{Mart{\'{\i}}nez-Delgado} et~al.}{2012}]{MartinezDelgado12}
{Mart{\'{\i}}nez-Delgado} D.,  et~al., 2012, \mn@doi [\apjl]
  {10.1088/2041-8205/748/2/L24}, \href
  {http://adsabs.harvard.edu/abs/2012ApJ...748L..24M} {748, L24}

\bibitem[\protect\citeauthoryear{{McConnachie}}{{McConnachie}}{2012}]{McConnachie12}
{McConnachie} A.~W.,  2012, \mn@doi [\aj] {10.1088/0004-6256/144/1/4}, \href
  {http://adsabs.harvard.edu/abs/2012AJ....144....4M} {144, 4}

\bibitem[\protect\citeauthoryear{{Moitinho}, {V{\'a}zquez}, {Carraro}, {Baume},
  {Giorgi}  \& {Lyra}}{{Moitinho} et~al.}{2006}]{Moitinho06}
{Moitinho} A.,  {V{\'a}zquez} R.~A.,  {Carraro} G.,  {Baume} G.,  {Giorgi}
  E.~E.,   {Lyra} W.,  2006, \mn@doi [\mnras]
  {10.1111/j.1745-3933.2006.00163.x}, \href
  {http://adsabs.harvard.edu/abs/2006MNRAS.368L..77M} {368, L77}

\bibitem[\protect\citeauthoryear{{Momany}, {Zaggia}, {Gilmore}, {Piotto},
  {Carraro}, {Bedin}  \& {de Angeli}}{{Momany} et~al.}{2006}]{Momany06}
{Momany} Y.,  {Zaggia} S.,  {Gilmore} G.,  {Piotto} G.,  {Carraro} G.,  {Bedin}
  L.~R.,   {de Angeli} F.,  2006, \mn@doi [\aap] {10.1051/0004-6361:20054081},
  \href {http://adsabs.harvard.edu/abs/2006A%26A...451..515M} {451, 515}

\bibitem[\protect\citeauthoryear{{More}, {Diemer}  \& {Kravtsov}}{{More}
  et~al.}{2015}]{More15}
{More} S.,  {Diemer} B.,   {Kravtsov} A.~V.,  2015, \mn@doi [\apj]
  {10.1088/0004-637X/810/1/36}, \href
  {http://adsabs.harvard.edu/abs/2015ApJ...810...36M} {810, 36}

\bibitem[\protect\citeauthoryear{{Moster}, {Naab}  \& {White}}{{Moster}
  et~al.}{2013}]{Moster13}
{Moster} B.~P.,  {Naab} T.,   {White} S.~D.~M.,  2013, \mn@doi [\mnras]
  {10.1093/mnras/sts261}, \href
  {http://adsabs.harvard.edu/abs/2013MNRAS.428.3121M} {428, 3121}

\bibitem[\protect\citeauthoryear{{Mouhcine}, {Ibata}  \& {Rejkuba}}{{Mouhcine}
  et~al.}{2010}]{Mouhcine10}
{Mouhcine} M.,  {Ibata} R.,   {Rejkuba} M.,  2010, \mn@doi [\apjl]
  {10.1088/2041-8205/714/1/L12}, \href
  {http://adsabs.harvard.edu/abs/2010ApJ...714L..12M} {714, L12}

\bibitem[\protect\citeauthoryear{{Nichols}, {Colless}, {Colless}  \&
  {Bland-Hawthorn}}{{Nichols} et~al.}{2011}]{Nichols11}
{Nichols} M.,  {Colless} J.,  {Colless} M.,   {Bland-Hawthorn} J.,  2011,
  \mn@doi [\apj] {10.1088/0004-637X/742/2/110}, \href
  {http://adsabs.harvard.edu/abs/2011ApJ...742..110N} {742, 110}

\bibitem[\protect\citeauthoryear{{Pe{\~n}arrubia}, {Benson}, {Walker},
  {Gilmore}, {McConnachie}  \& {Mayer}}{{Pe{\~n}arrubia}
  et~al.}{2010}]{Penarrubia10}
{Pe{\~n}arrubia} J.,  {Benson} A.~J.,  {Walker} M.~G.,  {Gilmore} G.,
  {McConnachie} A.~W.,   {Mayer} L.,  2010, \mn@doi [\mnras]
  {10.1111/j.1365-2966.2010.16762.x}, \href
  {http://adsabs.harvard.edu/abs/2010MNRAS.406.1290P} {406, 1290}

\bibitem[\protect\citeauthoryear{{Pe{\~n}arrubia}, {G{\'o}mez}, {Besla},
  {Erkal}  \& {Ma}}{{Pe{\~n}arrubia} et~al.}{2016}]{Penarrubia16}
{Pe{\~n}arrubia} J.,  {G{\'o}mez} F.~A.,  {Besla} G.,  {Erkal} D.,   {Ma}
  Y.-Z.,  2016, \mn@doi [\mnras] {10.1093/mnrasl/slv160}, \href
  {http://adsabs.harvard.edu/abs/2016MNRAS.456L..54P} {456, L54}

\bibitem[\protect\citeauthoryear{{Planck Collaboration} et~al.,}{{Planck
  Collaboration} et~al.}{2016}]{Planck16b}
{Planck Collaboration} et~al., 2016, preprint, \href
  {http://adsabs.harvard.edu/abs/2016arXiv160503507P} {} (\mn@eprint {arXiv}
  {1605.03507})

\bibitem[\protect\citeauthoryear{{Power}, {Wynn}, {Robotham}, {Lewis}  \&
  {Wilkinson}}{{Power} et~al.}{2014}]{Power14}
{Power} C.,  {Wynn} G.~A.,  {Robotham} A.~S.~G.,  {Lewis} G.~F.,   {Wilkinson}
  M.~I.,  2014, preprint, \href
  {http://adsabs.harvard.edu/abs/2014arXiv1406.7097P} {} (\mn@eprint {arXiv}
  {1406.7097})

\bibitem[\protect\citeauthoryear{{Read}, {Agertz}  \& {Collins}}{{Read}
  et~al.}{2016}]{Read16}
{Read} J.~I.,  {Agertz} O.,   {Collins} M.~L.~M.,  2016, \mn@doi [\mnras]
  {10.1093/mnras/stw713}, \href
  {http://adsabs.harvard.edu/abs/2016MNRAS.459.2573R} {459, 2573}

\bibitem[\protect\citeauthoryear{{Rich}, {Collins}, {Black}, {Longstaff},
  {Koch}, {Benson}  \& {Reitzel}}{{Rich} et~al.}{2012}]{Rich12}
{Rich} R.~M.,  {Collins} M.~L.~M.,  {Black} C.~M.,  {Longstaff} F.~A.,  {Koch}
  A.,  {Benson} A.,   {Reitzel} D.~B.,  2012, \mn@doi [\nat]
  {10.1038/nature10837}, \href
  {http://adsabs.harvard.edu/abs/2012Natur.482..192R} {482, 192}

\bibitem[\protect\citeauthoryear{{Rocha-Pinto}, {Majewski}, {Skrutskie},
  {Patterson}, {Nakanishi}, {Mu{\~n}oz}  \& {Sofue}}{{Rocha-Pinto}
  et~al.}{2006}]{RochaPinto06}
{Rocha-Pinto} H.~J.,  {Majewski} S.~R.,  {Skrutskie} M.~F.,  {Patterson} R.~J.,
   {Nakanishi} H.,  {Mu{\~n}oz} R.~R.,   {Sofue} Y.,  2006, \mn@doi [\apjl]
  {10.1086/503555}, \href {http://adsabs.harvard.edu/abs/2006ApJ...640L.147R}
  {640, L147}

\bibitem[\protect\citeauthoryear{{Roderick}, {Jerjen}, {Mackey}  \& {Da
  Costa}}{{Roderick} et~al.}{2015}]{Roderick15}
{Roderick} T.~A.,  {Jerjen} H.,  {Mackey} A.~D.,   {Da Costa} G.~S.,  2015,
  \mn@doi [\apj] {10.1088/0004-637X/804/2/134}, \href
  {http://adsabs.harvard.edu/abs/2015ApJ...804..134R} {804, 134}

\bibitem[\protect\citeauthoryear{{Romanowsky} et~al.,}{{Romanowsky}
  et~al.}{2016}]{Romanowsky16}
{Romanowsky} A.~J.,  et~al., 2016, \mn@doi [\mnras] {10.1093/mnrasl/slv207},
  \href {http://adsabs.harvard.edu/abs/2016MNRAS.457L.103R} {457, L103}

\bibitem[\protect\citeauthoryear{{Sales}, {Navarro}, {Cooper}, {White}, {Frenk}
   \& {Helmi}}{{Sales} et~al.}{2011}]{Sales11}
{Sales} L.~V.,  {Navarro} J.~F.,  {Cooper} A.~P.,  {White} S.~D.~M.,  {Frenk}
  C.~S.,   {Helmi} A.,  2011, \mn@doi [\mnras]
  {10.1111/j.1365-2966.2011.19514.x}, \href
  {http://adsabs.harvard.edu/abs/2011MNRAS.418..648S} {418, 648}

\bibitem[\protect\citeauthoryear{{Sales}, {Navarro}, {Kallivayalil}  \&
  {Frenk}}{{Sales} et~al.}{2016}]{Sales16}
{Sales} L.~V.,  {Navarro} J.~F.,  {Kallivayalil} N.,   {Frenk} C.~S.,  2016,
  preprint, \href {http://adsabs.harvard.edu/abs/2016arXiv160503574S} {}
  (\mn@eprint {arXiv} {1605.03574})

\bibitem[\protect\citeauthoryear{{Salvadori} \& {Ferrara}}{{Salvadori} \&
  {Ferrara}}{2009}]{Salvadori09}
{Salvadori} S.,  {Ferrara} A.,  2009, \mn@doi [\mnras]
  {10.1111/j.1745-3933.2009.00627.x}, \href
  {http://adsabs.harvard.edu/abs/2009MNRAS.395L...6S} {395, L6}

\bibitem[\protect\citeauthoryear{{Sand} et~al.,}{{Sand} et~al.}{2014}]{Sand14}
{Sand} D.~J.,  et~al., 2014, \mn@doi [\apjl] {10.1088/2041-8205/793/1/L7},
  \href {http://adsabs.harvard.edu/abs/2014ApJ...793L...7S} {793, L7}

\bibitem[\protect\citeauthoryear{{Sand}, {Spekkens}, {Crnojevi{\'c}}, {Hargis},
  {Willman}, {Strader}  \& {Grillmair}}{{Sand} et~al.}{2015}]{Sand15}
{Sand} D.~J.,  {Spekkens} K.,  {Crnojevi{\'c}} D.,  {Hargis} J.~R.,  {Willman}
  B.,  {Strader} J.,   {Grillmair} C.~J.,  2015, \mn@doi [\apjl]
  {10.1088/2041-8205/812/1/L13}, \href
  {http://adsabs.harvard.edu/abs/2015ApJ...812L..13S} {812, L13}

\bibitem[\protect\citeauthoryear{{Sawala} et~al.,}{{Sawala}
  et~al.}{2015}]{Sawala15}
{Sawala} T.,  et~al., 2015, \mn@doi [\mnras] {10.1093/mnras/stu2753}, \href
  {http://adsabs.harvard.edu/abs/2015MNRAS.448.2941S} {448, 2941}

\bibitem[\protect\citeauthoryear{{Sawala} et~al.,}{{Sawala}
  et~al.}{2016}]{Sawala16}
{Sawala} T.,  et~al., 2016, \mn@doi [\mnras] {10.1093/mnras/stv2597}, \href
  {http://adsabs.harvard.edu/abs/2016MNRAS.456...85S} {456, 85}

\bibitem[\protect\citeauthoryear{{Shattow} \& {Loeb}}{{Shattow} \&
  {Loeb}}{2009}]{Shattow09}
{Shattow} G.,  {Loeb} A.,  2009, \mn@doi [\mnras]
  {10.1111/j.1745-3933.2008.00573.x}, \href
  {http://adsabs.harvard.edu/abs/2009MNRAS.392L..21S} {392, L21}

\bibitem[\protect\citeauthoryear{{Simon} et~al.,}{{Simon}
  et~al.}{2015}]{Simon15}
{Simon} J.~D.,  et~al., 2015, \mn@doi [\apj] {10.1088/0004-637X/808/1/95},
  \href {http://adsabs.harvard.edu/abs/2015ApJ...808...95S} {808, 95}

\bibitem[\protect\citeauthoryear{{Simon} et~al.,}{{Simon}
  et~al.}{2016}]{Simon16}
{Simon} J.~D.,  et~al., 2016, preprint, \href
  {http://adsabs.harvard.edu/abs/2016arXiv161005301S} {} (\mn@eprint {arXiv}
  {1610.05301})

\bibitem[\protect\citeauthoryear{{Springel} et~al.,}{{Springel}
  et~al.}{2008}]{Springel08}
{Springel} V.,  et~al., 2008, \mn@doi [\mnras]
  {10.1111/j.1365-2966.2008.14066.x}, \href
  {http://adsabs.harvard.edu/abs/2008MNRAS.391.1685S} {391, 1685}

\bibitem[\protect\citeauthoryear{{Toloba}, {Guhathakurta}, {Romanowsky},
  {Brodie}, {Mart{\'{\i}}nez-Delgado}, {Arnold}, {Ramachandran}  \&
  {Theakanath}}{{Toloba} et~al.}{2016}]{Toloba16}
{Toloba} E.,  {Guhathakurta} P.,  {Romanowsky} A.~J.,  {Brodie} J.~P.,
  {Mart{\'{\i}}nez-Delgado} D.,  {Arnold} J.~A.,  {Ramachandran} N.,
  {Theakanath} K.,  2016, \mn@doi [\apj] {10.3847/0004-637X/824/1/35}, \href
  {http://adsabs.harvard.edu/abs/2016ApJ...824...35T} {824, 35}

\bibitem[\protect\citeauthoryear{{Torrealba}, {Koposov}, {Belokurov}  \&
  {Irwin}}{{Torrealba} et~al.}{2016a}]{Torrealba16b}
{Torrealba} G.,  {Koposov} S.~E.,  {Belokurov} V.,   {Irwin} M.,  2016a,
  \mn@doi [\mnras] {10.1093/mnras/stw733}, \href
  {http://adsabs.harvard.edu/abs/2016MNRAS.459.2370T} {459, 2370}

\bibitem[\protect\citeauthoryear{{Torrealba} et~al.,}{{Torrealba}
  et~al.}{2016b}]{Torrealba16a}
{Torrealba} G.,  et~al., 2016b, \mn@doi [\mnras] {10.1093/mnras/stw2051}, \href
  {http://adsabs.harvard.edu/abs/2016MNRAS.463..712T} {463, 712}

\bibitem[\protect\citeauthoryear{{Walker}, {Mateo}, {Olszewski}, {Bailey},
  {Koposov}, {Belokurov}  \& {Evans}}{{Walker} et~al.}{2015}]{Walker15}
{Walker} M.~G.,  {Mateo} M.,  {Olszewski} E.~W.,  {Bailey} III J.~I.,
  {Koposov} S.~E.,  {Belokurov} V.,   {Evans} N.~W.,  2015, \mn@doi [\apj]
  {10.1088/0004-637X/808/2/108}, \href
  {http://adsabs.harvard.edu/abs/2015ApJ...808..108W} {808, 108}

\bibitem[\protect\citeauthoryear{{Walker} et~al.,}{{Walker}
  et~al.}{2016}]{Walker16}
{Walker} M.~G.,  et~al., 2016, \mn@doi [\apj] {10.3847/0004-637X/819/1/53},
  \href {http://adsabs.harvard.edu/abs/2016ApJ...819...53W} {819, 53}

\bibitem[\protect\citeauthoryear{{Walsh}, {Willman}  \& {Jerjen}}{{Walsh}
  et~al.}{2009}]{Walsh09}
{Walsh} S.~M.,  {Willman} B.,   {Jerjen} H.,  2009, \mn@doi [\aj]
  {10.1088/0004-6256/137/1/450}, \href
  {http://adsabs.harvard.edu/abs/2009AJ....137..450W} {137, 450}

\bibitem[\protect\citeauthoryear{{Wetzel} \& {White}}{{Wetzel} \&
  {White}}{2010}]{Wetzel10}
{Wetzel} A.~R.,  {White} M.,  2010, \mn@doi [\mnras]
  {10.1111/j.1365-2966.2009.16191.x}, \href
  {http://adsabs.harvard.edu/abs/2010MNRAS.403.1072W} {403, 1072}

\bibitem[\protect\citeauthoryear{{Wetzel}, {Deason}  \&
  {Garrison-Kimmel}}{{Wetzel} et~al.}{2015}]{Wetzel15}
{Wetzel} A.~R.,  {Deason} A.~J.,   {Garrison-Kimmel} S.,  2015, \mn@doi [\apj]
  {10.1088/0004-637X/807/1/49}, \href
  {http://adsabs.harvard.edu/abs/2015ApJ...807...49W} {807, 49}

\bibitem[\protect\citeauthoryear{{Yanny} \& {Newberg}}{{Yanny} \&
  {Newberg}}{2016}]{Yanny16}
{Yanny} B.,  {Newberg} H.~J.,  2016, in {Newberg} H.~J.,  {Carlin} J.~L.,  eds,
   Astrophysics and Space Science Library Vol. 420, Astrophysics and Space
  Science Library. p.~63, \mn@doi{10.1007/978-3-319-19336-6_3}

\bibitem[\protect\citeauthoryear{{Yozin} \& {Bekki}}{{Yozin} \&
  {Bekki}}{2015}]{Yozin15}
{Yozin} C.,  {Bekki} K.,  2015, \mn@doi [\mnras] {10.1093/mnras/stv1828}, \href
  {http://adsabs.harvard.edu/abs/2015MNRAS.453.2302Y} {453, 2302}

\bibitem[\protect\citeauthoryear{{Zolotov} et~al.,}{{Zolotov}
  et~al.}{2012}]{Zolotov12}
{Zolotov} A.,  et~al., 2012, \mn@doi [\apj] {10.1088/0004-637X/761/1/71}, \href
  {http://adsabs.harvard.edu/abs/2012ApJ...761...71Z} {761, 71}

\bibitem[\protect\citeauthoryear{{de Jong}, {Butler}, {Rix}, {Dolphin}  \&
  {Mart{\'{\i}}nez-Delgado}}{{de Jong} et~al.}{2007}]{deJong07}
{de Jong} J.~T.~A.,  {Butler} D.~J.,  {Rix} H.~W.,  {Dolphin} A.~E.,
  {Mart{\'{\i}}nez-Delgado} D.,  2007, \mn@doi [\apj] {10.1086/517967}, \href
  {http://adsabs.harvard.edu/abs/2007ApJ...662..259D} {662, 259}

\bibitem[\protect\citeauthoryear{{van der Marel}, {Alves}, {Hardy}  \&
  {Suntzeff}}{{van der Marel} et~al.}{2002}]{vandermarel02}
{van der Marel} R.~P.,  {Alves} D.~R.,  {Hardy} E.,   {Suntzeff} N.~B.,  2002,
  \mn@doi [\aj] {10.1086/343775}, \href
  {http://adsabs.harvard.edu/abs/2002AJ....124.2639V} {124, 2639}

\makeatother
\end{thebibliography}
\label{lastpage}
\end{document}